# An inherently parallel $\mathcal{H}^2$-ULV factorization for solving dense linear systems on GPUs

## Qianxiang Ma[1]  and Rio Yokota[2]

## Abstract
Hierarchical low-rank approximation of dense matrices can reduce the complexity of their factorization from $\mathcal{O}(N^3)$ to $\mathcal{O}(N)$. However, the complex structure of such hierarchical matrices makes them difficult to parallelize. The block size and ranks can vary between the sub-blocks, which creates load imbalance. The dependency between the sub-blocks during factorization results in serialization. Since many sub-blocks are low-rank, their small computational load exposes the overhead of runtime systems. The combination of these factors makes it challenging to implement these methods on GPUs. In this work, we show that dense matrices can be factorized with linear complexity, while extracting the potential parallelism of GPUs. This is made possible through the $\mathcal{H}^2$-ULV factorization, which removes the dependency on trailing sub-matrices.

## 1. Introduction

Green's function matrices arising from boundary integral equations, covariance matrices in statistics, and kernel matrices in machine learning are large dense matrices, but their sub-blocks exhibit a low-rank structure. This structure can be exploited to reduce the complexity of multiplication and factorization of these matrices from $\mathcal{O}(N^3)$ to $\mathcal{O}(N)$. The accuracy can be controlled through a truncation threshold when compressing the low-rank blocks. Unlike sparse approximation of matrices, increasing the accuracy does not change the complexity, since the rank can become large but does not increase with $N$. Various structured low-rank matrix formats have been proposed such as the BLR matrix format from Amestoy et al. (2015), the BLR$^2$ matrix from Ashcraft et al. (2021), the HODLR format from Ambikasaran (2013), the HSS format from Chandrasekaran et al. (2006), and lastly $\mathcal{H}$-matrix and $\mathcal{H}^2$-matrices in Hackbusch (1999) and Hackbusch et al. (2000) correspondingly. BLR and BLR$^2$ are non-hierarchical structures, where the off-diagonal blocks of a block dense matrix are compressed to low-rank blocks. BLR$^2$ shares the low-rank basis between the blocks in each row and column, while BLR does not. HODLR and HSS are hierarchical structures, where the diagonal blocks are recursively subdivided. HSS shares the bases of the low-rank blocks among the rows and columns, while HODLR does not. $\mathcal{H}$ and $\mathcal{H}^2$ matrices are also hierarchical but subdivide not only the diagonal blocks but also the off-diagonal blocks. How far the blocks are subdivided is determined by the admissibility condition, which is based on a distance metric of the underlying geometry of the problem that forms the dense matrix. $\mathcal{H}^2$-matrices share the bases while $\mathcal{H}$-matrices do not. A subset of these structured low-rank matrix formats has been applied to matrix-vector and matrix-matrix products from Borm (2006), Cholesky and LU from Hackbusch (2015), QR factorizations in Ida et al. (2019), and eigenvalue decomposition in Benner and Mach (2012) and Ida (2022). Libraries that implement such algorithms can be used instead of BLAS and LAPACK when the matrix is large and has a low-rank structure.

The recent trend in hardware architecture is driven by machine learning, where low-precision matrix engines are becoming increasingly popular. Sparse matrices have difficulty extracting the full potential of such hardware since

[1]School of Computing, Tokyo Institute of Technology, Tokyo, Japan
[2]Global Scientific Information and Computing Center, Tokyo Institute of
Technology, Tokyo, Japan

**Corresponding author:**
Rio Yokota, Tokyo Institute of Technology, Rio Yokota Lab., 2-12-1 i7-
2 Ookayama Meguro-ku, Tokyo 152-8550, Japan.
Emails: ma@rio.gsic.titech.ac.jp, rioyokota@gsic.titech.ac.jp

these matrix engines are designed for dense matrices. Unlike sparse matrices, structured low-rank matrices consist of many small dense matrices, which can extract the potential of matrix engines through batched dense matrix operations. In addition, when using low-precision arithmetic, it does not make sense to perform the matrix operations exactly. Structured low-rank matrices can adjust their accuracy to match the precision of the matrix engines, and can reduce the complexity of matrix multiplication and factorization to $\mathcal{O}(N)$ while doing so. Therefore, we anticipate that there will be many cases where structured low-rank matrices can be used instead of BLAS/cuBLAS and LAPACK/cuSOLVER on future processor architectures with low-precision matrix engines. However, structured low-rank matrices are mainly studied by applied mathematicians and many implementations are written in Matlab. Even the few implementations on GPUs are for simple variants of structured low-rank matrices with suboptimal computational complexity and parallelism. It is true that structured low-rank matrices have a highly complex data structure and flow of computation, which makes them difficult to implement on GPUs. In the present work, we adopt the ULV factorization introduced from Chandrasekaran et al. (2006) to remove the dependency on trailing sub-matrices. This allows us to perform the factorization of an originally dense matrix in a highly parallel fashion, which allows us to extract the true potential of GPUs. Ma et al. (2022) extended the original ULV factorization for HSS matrices to $\mathcal{H}^2$-matrices, while preserving the highly parallel nature of the algorithm. However, their implementation only used CPUs, and the inherent parallelism of the ULV factorization has never been used to develop a highly parallel GPU implementation. Furthermore, Ma et al. only focus on the factorization, and the forward and backward substitution still has limited parallelism even when using their method. In the present work, we address these shortcomings of existing work and develop a highly scalable dense LU factorization method on GPUs. We also develop a novel algorithm that allows the forward and backward substitution to be performed in a highly parallel manner. Our main contributions can be summarized as follows.

- We develop for the first time a highly parallel GPU implementation of a dense LU factorization with linear complexity based on the $\mathcal{H}^2$-ULV factorization.
- We propose the very first inherently parallel forward and backward substitution algorithm for structured low-rank matrices.
- We provide an efficient implementation of our proposed algorithms of factorization and substitution that extracts the full potential of large GPU-accelerated systems.

We organize the rest of the paper as follows. The section "Background" contains the related work on other structured low-rank approximated matrix formats and existing attempts to make them run on parallel machines. In the section "Method", we describe our method as an augmentation to the existing ULV factorization for $H^2$-matrices. We cover aspects including the mathematical justification of pre-compressing Schur complements, and the modifications done to the construction, factorization, and substitution to make all of these phases inherently parallel. The section "Method" concludes with a complexity analysis of our proposed method. The section "Design Considerations for GPUs" describes the implementation and optimization details, along with the reasons why we have chosen to use specific libraries for GPUs and distributed memory implementation. We also provide details of our implementation and experiments for the sake of reproducibility. The section "Numerical Results" contains all the experiments we performed, including time complexity, FLOPS performance, relation between the rank and solution accuracy, strong scaling, and weak scaling results. Finally, we add closing remarks in the section "Conclusion."

## 2. Background

Our method belongs to a larger family of direct solvers for dense matrices that arise from kernel matrices generated in complex 3-D geometry problems. In this section, we compare our approach to other groups' notable work, that falls into the same or similar categories. We organize this section by the internal format of the structured low-rank approximation being used by different software packages and review them based on the resulting algorithmic complexity and the reported parallel capabilities.

### 2.1. Block (recursive block) factorization on matrices with independently low-rank approximated blocks

Independently approximated low-rank blocks applied to the dense matrices are closely related to the block-partitioned dense matrix. Therefore, many routines and methods designed for the dense matrices can be directly used on the approximated formats with slight modifications for low-rank representations. The block low-rank (BLR) matrix is the simplest of all structured low-rank approximated matrix formats and leads to the most simple implementations and parallelization. Although having a suboptimal $\mathcal{O}(N^2)$ factorization complexity and $\mathcal{O}(N^{1.5})$ in memory consumption, it is still much better than the dense $\mathcal{O}(N^3)$ complexity. HiCMA in Akbudak et al. (2017); Cao et al. (2019) and LORAPO in Cao et al. (2020, 2022) provide a highly optimized implementation in the BLR format that delivers excellent performance on CPUs and distributed memory environment. Many of the BLR implementations including the ones mentioned above use the 2-D block-cyclic process distribution used by ScaLAPACK (Blackford et al. 1997). The benefits come from (1) the workload is still balanced from the cyclic distribution after a row and a column have been eliminated and (2) both the communication volume and the neighbors in the process grid of each processor are $\sqrt{P}$, where $P$ is the total number of processors.

The multi-level approach of independently approximated blocks is named HODLR for weak admissibility and $\mathcal{H}$-matrix for strong admissibility. Despite having an even further reduction in factorization and memory complexity of $\mathcal{O}(N \log N)$, the $\mathcal{H}$-matrix format possesses a very complicated recursive structure of factorization data dependencies that is one of the most notorious problems to parallelize. A novel method has been presented by Chen and Martinsson (2022) for HODLR matrices that is capable of factorizing and applying forward/backward substitution in an entirely parallel fashion. Although their method is limited to the weak admissibility configuration, which shares a common weakness with the HSS matrices, they provide the first inherently parallel factorization to the families of the structured low-rank matrices with independently compressed blocks. More traditional approaches of recursive factorization algorithms for strongly admissible configurations are extremely difficult to parallelize, and the scalability of factorization using the H-matrix format is quite poor in the existing literature. One of the most notable distributed memory implementations of the $\mathcal{H}$-matrix $\mathcal{H}$ ACApK in and uses a modified structure named lattice $\mathcal{H}$-matrix, which embeds an $\mathcal{H}$-matrix structure within the dense blocks of a BLR matrix. The lattice $\mathcal{H}$-matrix combines the parallelism of the BLR matrix with the efficiency of the $\mathcal{H}$-matrix. Although good speed up on factorization has been reported by comparing with the BLR matrix formats, the algorithm struggles to scale for a relatively small number of cores used ($< 10^3$) due to the highly complex data-dependency structures. Unlike the HSS-ULV method, this approach suffers from the data dependency of trailing submatrices.

Our method differs significantly from approaches covered in this section of matrices compressed with independent bases. Even the simple flat structure of the BLR matrix possesses the factorization data dependency from the topleft corner to the bottom-right corner, and the reduction in algorithmic intensity via low-rank approximation has only reduced the parallel regions and exposed the data dependency even more. The methods described in the following section exploit the nature of the shared basis in order to avoid the data dependencies in the block and recursive block factorization algorithms.

## 2.2. Structured low-rank matrices with shared basis and ULV Factorization

Hierarchically semi-separable (HSS) matrices and the corresponding ULV factorization algorithms have first been introduced by Chandrasekaran et al. (2006) and had led to multiple parallel implementations (STRUMPACK: Rouet et al. (2016); H2pack: Huang et al. (2021)) that produced results that scale on extremely parallel machines on very large matrices (degrees of $10^8 \times 10^8$). Among them, STRUMPACK is a library that provides excellent support for running on NVIDIA GPUs and distributed memory configuration, and has made several significant contributions. H2pack is implemented on CPU only but uses a highly advanced construction method of HiDR introduced in Cai et al. (2022) that reports good accuracy and robustness for the low-rank approximation.

The ULV factorization method used weak admissibility structures, such as $BLR^2$ and HSS, which are inherently parallel by nature and have relatively simple structure. Before applying the internal LU or Cholesky factorization, ULV factorization methods first separate each block into the skeleton part and redundant part. When the shared row and column basis are factored out, this creates a sparsified matrix similar to that of the sparse multi-frontal LU solvers with nested dissection ordering. Therefore, the datadependency structure of the HSS-ULV factorization looks exactly like a multi-frontal LU solver with a constant dissector size (low-rank approximation rank) and has an entirely parallel region among each level with size $\mathcal{O}(2^l)$. The complexity of the method is also the same for a constant dissector size problem, which is strictly $\mathcal{O}(N)$. However, one weakness of the HSS-ULV factorization is its weak admissibility condition. When solving for kernel matrices in 3-D geometry, the off-diagonal low-rank blocks no longer compresses into a constant rank, but an $\mathcal{O}(N)$ rank that grows with the matrix dimension. The non-constant rank of the off-diagonal eventually makes the algorithm lose its linear complexity for the factorization and solution.

One of the notable attempts of running the ULV factorization for strongly admissible $\mathcal{H}^2$-matrices is made by Ma and Jiao (2018). They followed the same algorithms of the HSS-ULV factorization for the full basis transformations and block-LU factorization internally. With the addition of an innovative re-compression technique that has been widely used in factorization for independent basis formats, they can eliminate the fill-ins and maintain the linear complexity even for 3-D geometry. The inverse FMM and LoRaSp from Ambikasaran and Darve (2014) and Coulier et al. (2017) reported the same capability of factorizing strongly admissible $\mathcal{H}^2$-matrices also under linear complexity. LoRaSp uses extended sparsification to exclude the low-rank blocks from the dense operations, as well as extending the factorization algorithm to an initially sparse matrix and uses low-rank approximation hierarchically to compress the fill-ins. Both ULV factorization and LoRaSp rely on the creation of an intermediate block sparse representation while recompressing the fill-ins to maintain the $\mathcal{O}(N)$ algorithmic complexity. However, even though the approaches by Ma and Jiao (2018), Ambikasaran and Darve (2014), and Coulier et al. (2017) are able to extend the HSS-ULV factorization to 3-D problems while maintaining the linear complexity, they are not able to inherit the highly parallel nature of the HSS-ULV factorization because the recompression brings back the dependency for the trailing submatrices. Distributed memory parallelization attempts for the strongly admissible $\mathcal{H}^2$-matrices have been made by LoRaSp, but the reported parallel scalability is far from that of the HSS-matrices.

The work by Ma et al. (2022) has been the first attempt to unite the advantages of parallel capabilities of the ULV factorization to the ideal linear complexity from the strong admissibility configurations of $\mathcal{H}^2$-matrices. The results presented include a highly parallel CPU implementation both in shared memory as well as distributed memory configuration that shows good strong scaling up to 10,000 CPU cores. Although their method can potentially be extended to GPUs, such an effort has been left for future work. Also left for future work was the forward and backward substitution phase, which was not implemented in parallel in their work. With traditional factorization methods for structured low-rank matrices, it has been difficult to extract the full potential of GPUs. This is because the arithmetic intensity of the low-rank blocks is quite low, and the dependency between the blocks prevents them from being processed in batches. The present work exploits the lack of dependency in the $\mathcal{H}^2$-ULV method by Ma et al. (2022) and shows that a scalable batched GPU implementation is indeed possible for this algorithm. Further, we supplement the approach taken by Ma et al. (2022) with a more accurate and sound mathematical proof. We also extend their attempt of parallelization into an inherently parallel forward and backward substitution that accompanies the $\mathcal{H}^2$-ULV factorization. To the best of our knowledge, this is the first work that proposes an inherently parallel forward and backward substitution for dense matrices.

# 3. Method

The main difference of our work from the other $\mathcal{H}^2$-matrix factorization methods is the introduction of another set of bases, which we called the factorization basis. The factorization basis, in short, is a low-rank shared basis that approximates all of the Schur complements computation that will arise during the factorization, even if the Schur complement has been computed from the closely interacting matrix (in other words, dense matrix) blocks. The introduction of the factorization basis is to augment the ULV factorization method that is initially used for factorizing and inverting shared-basis structured low-rank matrices, such as BLR$^2$, HSS, and $\mathcal{H}^2$ matrices. Merging the factorization basis with the shared basis for approximating the low-rank matrix blocks removes the need to update the basis during the recompression of the fill-in blocks. This results in an inherently parallel LU factorization even for dense matrices arising from 3-D geometry.

We first clarify what we mean by "compression of the Schur complement," where we are addressing the compression of a particular matrix block $C_{jk} = A_{ji}A_{ii}^{-1}A_{ik}$, which is, in fact, the difference matrix between the Schur complement and the original matrix block. The proper definition of Schur complement is given in the following, which is a full-rank matrix block located on the diagonal:

$$M/A_{jj} = A_{jj} - A_{ji}A_{ii}^{-1}A_{ij} \tag{1}$$

Compressing only the Schur complement update uses significantly lower numeric rank even for very closely interacting matrix blocks, including completely overlapping geometry such as for updating matrix locating on the diagonal. The low-rank approximation rank correlates very weakly with the size of the input matrix, behaving more similarly to the compression of the far-field interactions, instead of having a rank that grows with $\mathcal{O}(N)$ for 3-D problems. We provide a comparison of the rank and difference in the accuracy of our method versus the traditional means of compressing the closely interacting off-diagonal blocks that the HODLR and HSS matrices use later in the results section.

## 3.1. Characteristics of the factorization basis

The primary functionality of introducing the factorization basis is to approximate the difference matrix in Schur complement with the original matrix, given as

$$A_{ji}A_{ii}^{-1}A_{ik} \approx U_j \Sigma V_k^T, \forall i \neq j \& i \neq k \tag{2}$$

$\mathcal{H}^2$-matrix factorization methods proposed by Ma and Jiao (2018), Ambikasaran and Darve (2014), and Coulier et al. (2017) also use a factorization basis to recompress the fill-in blocks. The significance of the method proposed by Ma et al. (2022) is that the factorization basis is pre-computed and merged into the shared basis before the factorization. This way, the shared basis once constructed in the beginning is not modified later during the recompression of the blocks that fill-in (low-rank blocks that become dense) during the factorization. Even with trailing updates introduced, the approximation with the same basis still holds as the Woodbury matrix identity of matrix inversion suggests that

$$\left(A_{ii} - A_{il}A_{ll}^{-1}A_{li}\right)^{-1} = A_{ii}^{-1} + A_{ii}^{-1}A_{il}\dots A_{li}A_{ii}^{-1} \quad (3)$$

Multiplying with $A_{ji}$ and $A_{ik}$ from the sides, equation (3) becomes

$$A_{ji}\left(A_{ii}^{-1} + A_{ii}^{-1}A_{il}\dots A_{li}A_{ii}^{-1}\right)A_{ik} \approx U_j(\Sigma + \Sigma_1\dots\Sigma_2)V_k^T \quad (4)$$

Suggesting that the Schur complement update produced from the updated diagonal is still compressible using the same basis.

The updates introduced to the off-diagonal blocks can also be compressed using the same logic:

$$\left(A_{ji} - A_{jl}A_{ll}^{-1}A_{li}\right)A_{ii}^{-1}A_{ik} \approx U_j(\Sigma - \Sigma_1\dots\Sigma_2)V_k^T \quad (5)$$

The purpose of constructing the factorization basis is to prevent the need to update the shared basis during the ULV factorization. Therefore, we want all possible fill-ins that might occur during the ULV factorization to occur during the construction of the factorization basis. Contrary to sparse matrix factorization where we would like to minimize the amount of fill-in, here we want to maximize the fill-in so that the factorization basis includes all possible fill-ins. Therefore, as we construct the factorization basis, we not only consider the updates from the ordinary elimination order from top-left to bottom-right, but also the updates that come from the latter block eliminations. This is also the key to making this phase inherently parallel. Because we are considering the elimination of arbitrary order, we are free to perform permutations in the factorization ordering during the computation of the factorization basis. This means that we can compute the fill-ins for each block row/column independently, so this operation is inherently parallel.

## 3.2. Trailing submatrix update for single level

This subsection describes the trailing submatrix updates for a single-level factorization. For all structured low-rank matrices with shared bases (HSS, $\mathcal{H}^2$-matrices, etc.), the ULV-based factorization methods progress on a level-by-level order that allows us to deal with a simpler single-level variant of the problem before recursing on to the next.

We show how the factorization basis takes advantage of the low-rankness of the Schur complement updates during the factorization, to enable an inherently parallel factorization algorithm even when the matrix contains many dense off-diagonal blocks.

By the term "skeleton," we address the matrix contents that correspond to the compressible or projected region to the basis as the U and V transformations are applied from the sides of the $\mathcal{H}^2$-matrix. In contrast, the regions that are dropped from the basis projection are named "redundant." During the ULV factorization of the $\mathcal{H}^2$-matrix, the factorization basis redirects trailing updates to only update the matrix blocks corresponding to the skeleton so that the redundant matrix blocks are never updated until they are eliminated from the factorization. This absence of trailing updates leads to an inherently parallel factorization. In the remainder of the paper, the dense blocks that correspond to the different regions of the basis, are named $A^{SS}$, $A^{SR}$, $A^{SR}$, and $A^{RR}$, to indicate the contents have been obtained from either the skeleton or the redundant regions of the left or right basis, as shown in Figure 2. The concept of a factorization basis is independent of the traditional low-rank shared basis and can be used on matrix structures that do not have initial low-rank approximations, such as a fully block-dense matrix or a sparse matrix. Coulier et al. (2017) use this concept to form a sparse direct solver that recompresses the fill-ins using the factorization basis.

In Figure 1, we show the first step of an LU factorization of a symmetric block-dense matrix, where the first block row/column has been factorized. For simplicity, we will use a $3 \times 3$ block dense matrix to introduce the basic concepts, before moving to the method. In our method, the Schur complement update is presumed to be numerically low-rank, but Figure 1 shows the updates to be dense. In this case, it is obvious that the factorization of trailing submatrices (shown in grey) cannot proceed until the Schur complement updates are computed.

We now introduce the concept of low-rank approximation, where the SVD of a submatrix $A_{jk}$ can be written as

$$A_{jk} = \begin{pmatrix} U_j^S & U_j^R \end{pmatrix} \begin{pmatrix} A_{jk}^{SS} & A_{jk}^{SR} \\ A_{jk}^{RS} & A_{jk}^{RR} \end{pmatrix} \begin{pmatrix} V_k^R & V_k^R \end{pmatrix}^T \quad (6)$$

As mentioned earlier, we introduce the concept of the "skeleton" part (with superscript $S$) and the "redundant" part (with superscript $R$). For a low-rank block $U^R$, $V^R$, $A^{RS}$, $A^{SR}$, and $A^{RR}$ are zero, but for a dense matrix they are non-zero. This allows us to partition both the dense and low-rank blocks using the same split between the skeleton and redundant parts. We then permute the skeleton and redundant parts for all blocks:

$$A_{jk} = \begin{pmatrix} U_j^R & U_j^S \end{pmatrix} \begin{pmatrix} A_{jk}^{RR} & A_{jk}^{RS} \\ A_{jk}^{SR} & A_{jk}^{SS} \end{pmatrix} \begin{pmatrix} V_k^R & V_k^S \end{pmatrix}^T \quad (7)$$

This results in a conversion from a standard $\mathcal{H}^2$-matrix to an $\mathcal{H}^2$-ULV matrix as shown in Figure 2. The process of operating on each dense or low-rank block is included in Figure 2, as well as the global view where the original matrix is a mixture of dense and low-rank blocks.

In Figure 3, we show how the elimination of the first block row/column results in fill-ins for the $A^{SS}$ part only. Note that this holds even when no low-rank blocks exist, and only requires the Schur complement updates to be

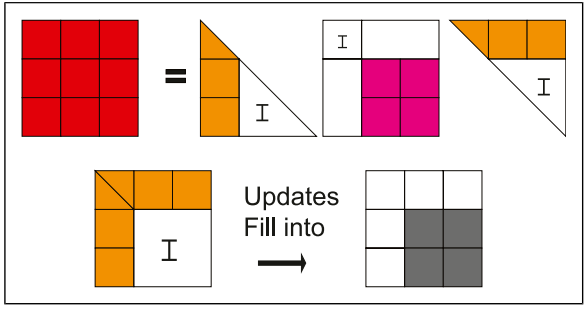

**Figure 1.** Step 1 of LU factorization on a 3 × 3 dense matrix with trailing Schur complement updates.

low-rank. This is more clear when expressed in the form of an equation:

$$\left(U_j^T A_{ji} U_i\right)\left(U_i^T A_{ii}^{-1} U_i\right)\left(U_i^T A_{ik} U_k\right) \approx \begin{pmatrix} 0 & 0 \\ 0 & A^{SS} \end{pmatrix} \quad (8)$$

Other trailing block partitions $A^{RR}$, $A^{RS}$, and $A^{SR}$ are not being updated from the defined behavior in equation (8). Transforming this equation by separating it into different parts of $R$ and $S$, we can obtain the following list of equations that are easier to comprehend:

$$\begin{cases} A_{ji}^{RX}\left(A_{ii}^{XX}\right)^{-1} A_{ik}^{XR} = 0 \\ A_{ji}^{RX}\left(A_{ii}^{XX}\right)^{-1} A_{ik}^{XS} = 0 \\ A_{ji}^{SX}\left(A_{ii}^{XX}\right)^{-1} A_{ik}^{XR} = 0 \\ A_{ji}^{SX}\left(A_{ii}^{XX}\right)^{-1} A_{ik}^{XS} \neq 0 \\ X := R \| S \end{cases} \quad (9)$$

Figure 4 shows a complete view of the entire factorization, where the invariant matrix blocks with no update data dependency are shown in green and the updated matrix blocks with data dependency are shown in pink. The left figure shows the original view where each block is partitioned into the skeleton and redundant parts. The right figure shows the permuted view where the skeleton part is clustered to the upper-left. From this view, we can see that all of the possible locations of updates are clustered to the bottom-right corner of the $A^{SS}$ region and have removed the trailing diagonal update dependencies inside the $A^{RR}$ region, enabling a parallel elimination of all blocks in different rows and columns. Compared with the traditional approach for a block LU factorization on dense matrices, we have exposed a much larger parallel region by taking advantage of the low-rank compressible Schur complement updates. This concept only requires the Schur complement updates to be low-rank, and remains inherently parallel regardless of the number of dense off-diagonal blocks.

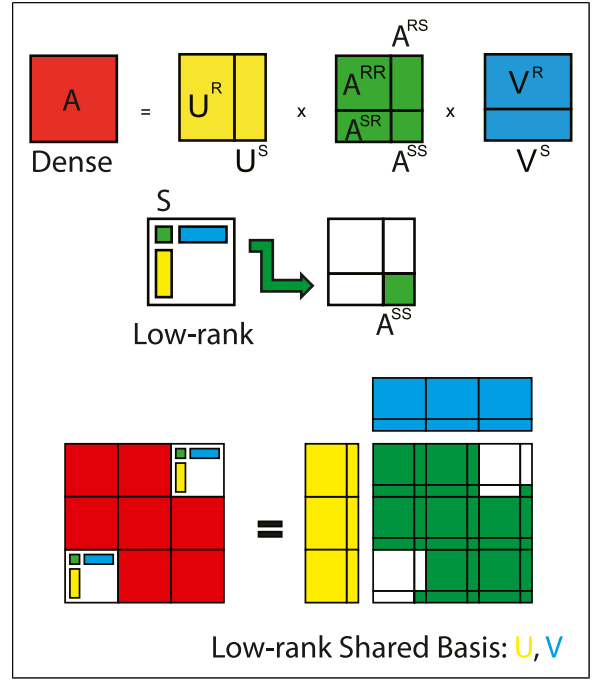

**Figure 2.** Matrix sparsification from full-size shared basis.

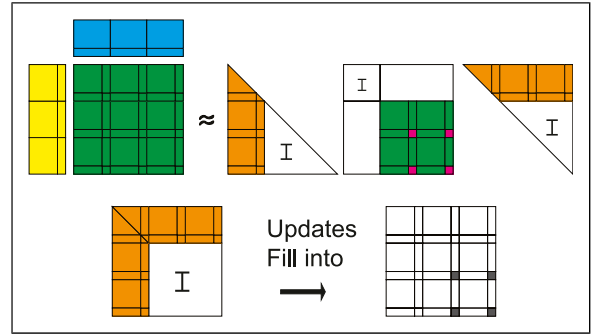

**Figure 3.** Step 1 of the internal block-LU factorization on a 3 × 3 dense matrix during ULV factorization, and corresponding update locations.

### 3.3. Trailing submatrix update for multi-level

For multi-level ULV factorization of strongly admissible $\mathcal{H}^2$-matrices, we must deal with the trailing submatrix update patterns for the $A^{SS}$ blocks across levels. Filling into the blocks, which are originally low-rank, leads to extra dense matrix blocks created for the next level and eventually having super linear factorization complexity. The chain of fill-ins is a critical problem reported by the original $\mathcal{H}^2$-ULV factorization and LoRaSp. This is the primary reason for introducing the re-compression of the fill-in blocks. In this section, we focus exclusively on the updates that go to the next level of factorization by looking at a simple 3-D cubic geometry which requires a strong admissibility $\mathcal{H}^2$-matrix as shown in Figure 5. In this example, the interactions

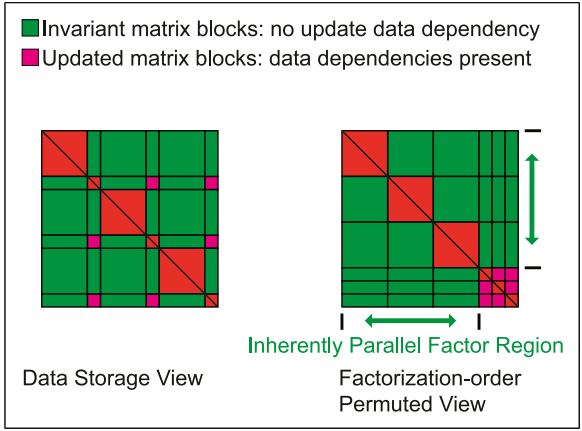

■ Invariant matrix blocks: no update data dependency
■ Updated matrix blocks: data dependencies present

Inherently Parallel Factor Region

Data Storage View

Factorization-order
Permuted View

**Figure 4.** Any symmetric permutation of elimination order, invariant matrix blocks are not updated until elimination → Permute to eliminate all $A^{RR}$ before $A^{SS}$ to form an inherently parallel factoring region.

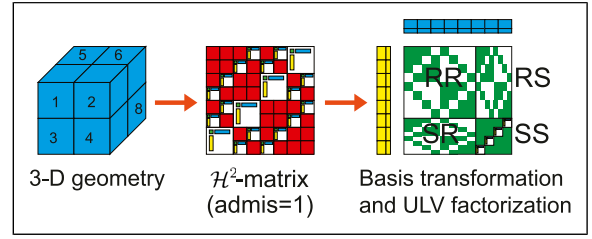

3-D geometry

$\mathcal{H}^2$-matrix (admis=1)

Basis transformation and ULV factorization between cubes that share the same surface are considered as dense blocks. Clustering the redundant part to the upper-left, results in a sparsity pattern shown in the rightmost figure. To consider the sparsity pattern at the next level, we zoom in to take a closer look at the updates created for the $A^{SS}$ region.

**Figure 5.** A $\mathcal{H}^2$-matrix compression to a simple 3-D geometry.

On the top picture in Figure 6, we show how directly multiplying the factorized $A^{SR}$ with $A^{RS}$ creates a substantial amount of updates. Some dense matrix blocks are filled into the upper-level low-rank matrices and converted to dense matrix blocks based on the sparsity pattern alone. The naively computed fill-ins do not exploit the low-rankness of the factorization basis to its full extent. In the following analysis, we show that the actual pattern of updates can be limited to only the diagonal blocks of the $A^{SS}$, as shown in the bottom picture in Figure 6.

To be consistent with the notation used, we again use the letter combination $C_{ij}$ to denote the Schur complement update being introduced to the superscript $SS$ region of the $(i, j)$ matrix block. Similar to the updates in the regions correlating to the redundant basis, we utilize the low-rank approximation results and use the factorization basis for redirecting the updated locations, to expose sparsity (zero entries) in the update matrix. This property of the factorization basis removes the need for re-compression during the ULV factorization and greatly simplifies the overall implementation.

We divide the proof into three incremental steps that correlates to the different means for the trailing panel updates produced. Firstly, we show that for all updates introduced to the diagonal blocks:

$$C_{jj} = A_{ji}^{SR} \left( A_{ii}^{RR} \right)^{-1} A_{ij}^{RS} = 0, \forall i \neq j \tag{10}$$

Secondly, we show that for all updates introduced to the off-diagonal blocks:

$$C_{jk} = A_{ji}^{SR} \left( A_{ii}^{RR} \right)^{-1} A_{ik}^{RS} = 0, \forall i \neq j \neq k \tag{11}$$

Lastly, we show that for a single $A^{RS}$ or $A^{SR}$ that lies in the same block as the eliminated $A^{RR}$:

$$\begin{cases} C_{ji} = A_{ji}^{SR} \left( A_{ii}^{RR} \right)^{-1} A_{ii}^{RS} = 0, \forall i \neq j \\ C_{ij} = A_{ii}^{SR} \left( A_{ii}^{RR} \right)^{-1} A_{ij}^{RS} = 0, \forall i \neq j \end{cases} \tag{12}$$

We can prove equation (10) by looking at the following inversion of the $2 \times 2$ block dense linear system, taken from portions the $\mathcal{H}^2$-matrix after the transformations of $U$ and $V$:

$$\begin{pmatrix} A_{jj}^{SS} & A_{jj}^{SR} \\ A_{jj}^{RS} & A_{jj}^{RR} \end{pmatrix}^{-1} = \begin{pmatrix} p & r \\ q & s \end{pmatrix} \tag{13}$$

We assume that the diagonal blocks $A_{jj}^{SS}$ and $A_{ii}^{RR}$ are invertible, and that $A_{ji}^{SR}$ and $A_{ij}^{RS}$ are non-zero matrices (otherwise $C_{ij} = 0$ is trivial). We know that the elimination of block $A_{jj}$ produces zero updates to $A_{ii}^{RR}$, and $A_{ii}^{SS}$ being a subsystem contained inside $A_{jj}$. The updates in $A_{ii}^{RR}$ from the elimination of $A_{jj}^{SS}$ is also a zero matrix $A_{ij}^{RS}(A_{jj}^{SS})^{-1}A_{ji}^{SR} = 0$, and the inverted system can be written as

$$\begin{pmatrix} \left( A_{jj}^{SS} - C_{jj} \right)^{-1} & - \left( A_{jj}^{SS} \right)^{-1} A_{ji}^{SR} \left( A_{ii}^{RR} \right)^{-1} \\ - \left( A_{ii}^{RR} \right)^{-1} A_{ij}^{RS} \left( A_{jj}^{SS} \right)^{-1} & \left( A_{ii}^{RR} \right)^{-1} \end{pmatrix} \tag{14}$$

Thus, multiplying the original system with the inverted form should result in an identity matrix:

$$\begin{pmatrix} A_{jj}^{SS} p + A_{ji}^{SR} q & A_{jj}^{SS} r + A_{ji}^{SR} s \\ A_{ij}^{RS} p + A_{ii}^{RR} q & A_{ij}^{RS} r + A_{ii}^{RR} s \end{pmatrix} = \begin{pmatrix} I & 0 \\ 0 & I \end{pmatrix} \tag{15}$$

Expanding the lower-left block on both sides, we have

$$A_{ij}^{RS} \left( A_{jj}^{SS} - C_{jj} \right)^{-1} - A_{ii}^{RR} \left( A_{ii}^{RR} \right)^{-1} A_{ij}^{RS} \left( A_{jj}^{SS} \right)^{-1} = 0 \tag{16}$$

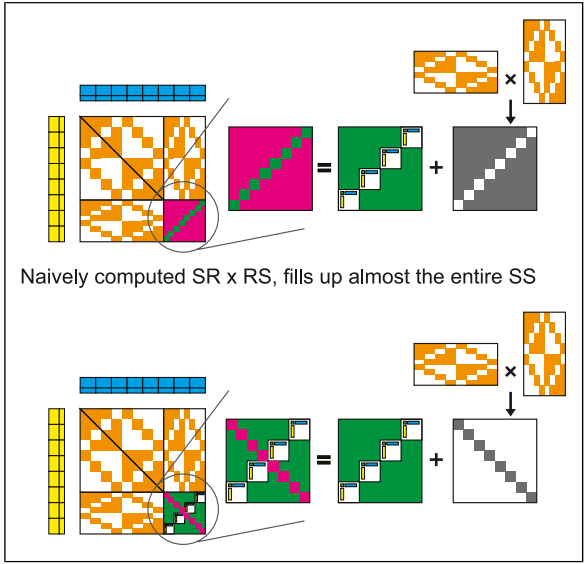

**Figure 6.** Acutal updated locations for ULV factorization when eliminating $A^{RR}$, symmetric to $A^{SS}$ elimination.

Given that all blocks are non-zero matrices, we complete the proof of $C_{jj} = 0$ in equation (10).

The proof for equations (11) and (12) can be arranged together by looking at a $3 \times 3$ system and extending the proof of equation (10)

$$A' = \begin{pmatrix} A_{ii}^{RR} & A_{ij}^{RS} & A_{ik}^{RS} \\ A_{ji}^{SR} & A_{jj}^{SS} & A_{jk}^{SS} \\ A_{ki}^{SR} & A_{kj}^{SS} & A_{kk}^{SS} \end{pmatrix} \quad (17)$$

The proof for the two cases focuses on the updates introduced to block $A_{jk}^{SS}$ and $A_{kj}^{SS}$, from the elimination of $A_{ii}^{RR}$. The sole difference lies in whether $k = i$ or $k \neq i$, affecting whether the updates introduced to the last block of $A_{kk}^{SS}$ are zero matrices or not. We can proceed with the proof since this difference in the trailing update characteristic does not play a significant role. Considering another similar system to the previous $A'$ system by applying permutations to the 1st and 2nd block rows and columns we have

$$A'' = \begin{pmatrix} & & I \\ & I & \\ I & & \end{pmatrix} A' P^T = \begin{pmatrix} A_{jj}^{SS} & A_{ji}^{SR} & A_{jk}^{SS} \\ A_{ij}^{RS} & A_{ii}^{RR} & A_{ik}^{RS} \\ A_{kj}^{SS} & A_{ki}^{SR} & A_{kk}^{SS} \end{pmatrix} \quad (18)$$

Since the applied permutation keeps the bottom right block $A_{33}'' = A_{33}'$, this equality also holds when creating the inverse of both systems. Writing out the equations for both

$$\begin{cases} C_{kk}' = \left( A_{kj}^{SS} + C_{kj} \right) \left( A_{jj}^{SS} \right)^{-1} \left( A_{jk}^{SS} + C_{jk} \right) \\ C_{kk}'' = A_{kj}^{SS} \left( A_{jj}^{SS} \right)^{-1} A_{jk}^{SS} \\ \left( A_{33}' \right)^{-1} = \left( A_{kk}^{SS} - A_{ki}^{SR} \left( A_{ii}^{RR} \right)^{-1} A_{ik}^{RS} - C_{kk}' \right)^{-1} \\ \left( A_{33}'' \right)^{-1} = \left( A_{kk}^{SS} - C_{kk}'' - A_{ki}^{SR} \left( A_{ii}^{RR} \right)^{-1} A_{ik}^{RS} \right)^{-1} \end{cases} \quad (19)$$

For obtaining these equations, we consider the block LU factorization applied to the first and second rows correspondingly and let them update the bottom right block. For the $A'$ system, we assumed that the elimination of $A_{ii}^{RR}$ introduces updates to all trailing blocks except $A_{jj}^{SS}$, which is already proven by equation (10). The elimination of $A_{jj}^{SS}$ in the $A'$ system accommodates the updates introduced and carries it over to the bottom right block. For the $A''$ system, the elimination of $A_{jj}^{SS}$ updates only the $A_{kk}^{SS}$, as the other blocks lie in the regions of redundancy and by using the factorization basis they do not require any updates. The following elimination of the block $A_{ii}^{RR}$ operates on its original contents to produce the trailing update. By comparing these relations, we can identify that $\left( A_{33}' \right)^{-1} = \left( A_{33}'' \right)^{-1}$ implies $C_{kk}' = C_{kk}''$. We can formulate the result of $C_{kk}' - C_{kk}''$ as the following:

$$C_{kk}' - C_{kk}'' = \begin{pmatrix} A_{kj}^{SS} \\ C_{kj} \end{pmatrix}^T \begin{pmatrix} 0 & \left( A_{jj}^{SS} \right)^{-1} \\ \left( A_{jj}^{SS} \right)^{-1} & \left( A_{jj}^{SS} \right)^{-1} \end{pmatrix} \begin{pmatrix} A_{jk}^{SS} \\ C_{jk} \end{pmatrix} \quad (20)$$

No other $C_{kj}$ and $C_{jk}$ combinations other than $C_{kj} = C_{jk} = 0$ can make $C_{kk}' - C_{kk}'' = 0$ hold for arbitrary non-zero entries. Thus, we complete the proof for both equation (11) and (12), by setting $k \neq i$ and $k = i$ correspondingly.

For the remaining case where $i = j = k$, we made a generic assumption that the matrix is non-singular. The factorization of the redundant matrix will produce a non-zero update to the skeleton matrix of the same origin matrix before the orthogonal transformations are applied. In other words, $C_{ii} = A_{ii}^{SR} (A_{ii}^{RR})^{-1} A_{ii}^{RS} \neq 0$. Combining the proof for equations (10)–(12) on the updates introduced to regions of $A^{SS}$ gives

$$A_{ji}^{SR} \left( A_{ii}^{RR} \right)^{-1} A_{ik}^{RS} \begin{cases} \neq 0, \forall i = j = k \\ = 0, \forall i \neq j \| i \neq k \end{cases} \quad (21)$$

Equation (21) guarantees the number of dense matrix blocks to be bounded to the same structure as it is being constructed. Therefore, the only computed updates lie on the diagonal blocks.

Briefly summarizing this section, unlike the original ULV factorization or the inverse FMM based on extended sparsification, our method of using factorization basis by design does not require the usage of recompression during factorization. Using the factorization basis and pre-compressing the trailing updates can be seen as a means to both introduce parallelism by eliminating data dependencies, and also maintain linear algorithmic complexity. These properties play a major role in the implementation of the parallel ULV factorization and substitution, which are described in the later sections.

### 3.4. $\mathcal{H}^2$-matrix construction

The construction of the $\mathcal{H}^2$-matrix includes steps to build the low-rank shared bases, and forming the low-rank coupling matrix for each individual low-rank block. The low-rank approximations applied to each row or column, are independent of each other and can be executed in an embarrassingly parallel fashion. Since the focus of this paper is on the parallelization of the matrix factorization and substitution, we do not go into the details of the $\mathcal{H}^2$-matrix construction nor provide results corresponding to this section.

In this work, we adopt a conventional method for constructing the shared basis known as interpolative decomposition (ID) as shown in Figure 7. The key difference between our method and conventional $\mathcal{H}^2$-matrix construction methods is that we include a factorization basis along with the low-rank approximation basis, which introduces some extra work and requires a pre-factorization step. As shown in Figure 7, ID is based on the sampling of points in the underlying geometry. More complicated and better-performing sampling methods such as HiDR used by H2pack or GOFMM can approximate to the same degree or better accuracy using even fewer number of sampled bodies, but their method is beyond the scope of this paper so we do not discuss it further. Disabling the sampling and always using the entire domain (which is $\mathcal{O}(N)$ size) of well-separated points will lead to a $\mathcal{O}(N^2)$ construction cost but having the best construction accuracy, while any constant sample size reduces this complexity to $\mathcal{O}(N)$.

In order to include the factorization basis of compressed Schur complement updates in equation (2), we combine the contents of the factorization basis of the near-field with the low-rank basis of the far-field, as shown in Figure 8. Instead of forming all of the possible combinations $i$, $j$, $k$ in $A_{ji}A_{ii}^{-1}A_{ik}$ of the Schur complement that arise during the factorization, a simpler alternative is to approximate the term $A_{ji}A_{ii}^{-1}$ instead for approximating $U_j$. The expression $A_{ji}A_{ii}^{-1}$ uses the same row basis and poses an upper bound in rank for all possible $k$ in different Schur complement $A_{ji}A_{ii}^{-1}A_{ik}$:

$$A_{ji}A_{ii}^{-1}A_{ik} \approx \begin{pmatrix} U_j^R & U_j^S \end{pmatrix} \begin{pmatrix} 0 & 0 \\ 0 & A^{SS} \end{pmatrix} \begin{pmatrix} U_k^R & U_k^S \end{pmatrix}^T \quad (22)$$

In order to produce the internal zero matrix blocks on the first row of the $A^{SS}$ matrix, we need the first two terms on the left hand side to satisfy

$$A_{ji}A_{ii}^{-1} \approx \begin{pmatrix} U_j^R & U_j^S \end{pmatrix} \begin{pmatrix} 0 & 0 \\ p & q \end{pmatrix} \quad (23)$$

Working backwards, $A_{ji}A_{ii}^{-1}$ needs to be compressible when using the $U_j$ factorization basis, if $U_j$ can approximate the Schur complement $A_{ji}A_{ii}^{-1}A_{ik}$. By combining the contents of the far field with the $A_{ji}A_{ii}^{-1}$ block and applying a low-rank approximation, we can obtain a composite basis that serves as both the factorization basis and low-rank shared basis.

### 3.5. Pre-factorization

In this section, we focus more on the process of obtaining the compressible Schur complement instead of the means of applying a low-rank approximation on it, for the reason that the compressible contents can be concatenated together with the originally low-rank components of the $\mathcal{H}^2$-matrix to form a composite basis in the construction phase. The difference in the different methods of low-rank approximation can have a large difference in accuracy and performance according to different pieces of literature (Borm, 2006; Cai et al., 2022).

In order to include the factorization basis into the shared low-rank basis, we essentially need to factorize the matrix twice; one for the pre-factorization to compute the factorization basis, and another to perform the actual ULV factorization using this shared basis. The fact that both factorizations are inherently parallel still gives us an advantage over existing methods, but we would like to reduce this overhead if possible. In order to alleviate the cost of factorizing the matrix twice, we pre-process and approximate the joint term of $A_{ji}A_{ii}^{-1}$ before the actual low-rank approximation for constructing the basis. One approach is to also adopt sampling for the near-field boxes, as shown in Figure 8. Sampling of the near-field does not reduce the complexity like the sampling for the far-field since the leaf block size is bounded by a constant anyway, but it does reduce the overhead of factorizing twice by a constant. This constant can actually be quite large, since factorization of a dense matrix has $\mathcal{O}(N^3)$ complexity.

Furthermore, we discovered that without factorizing the close field interaction $A_{ii}$, a good approximation on the content and the rank of $A_{ji}A_{ii}^{-1}$ can be found via iterative solvers. In our implementation which internally uses Cholesky factorization for the dense diagonals, we assumed the kernel function defined generates an input matrix with semi-positive definiteness. Therefore, we used the Gauss-Seidel iterative method for approximating the contents of $A_{ji}A_{ii}^{-1}$ without explicitly factorizing it. For the kernel functions that we have tested, one or two Gauss-Seidel iterations produce a sufficiently accurate approximation of $A_{ji}A_{ii}^{-1}$. We close the description of the construction phase by providing the complete algorithm of $\mathcal{H}^2$-matrix construction using interpolative decomposition as an example while including the pre-factorization, in Algorithm 1. The Gauss-Seidel optimization is specific to the step where $A_{c,c}^{-1}$ is computed in 11 in the provided algorithm.

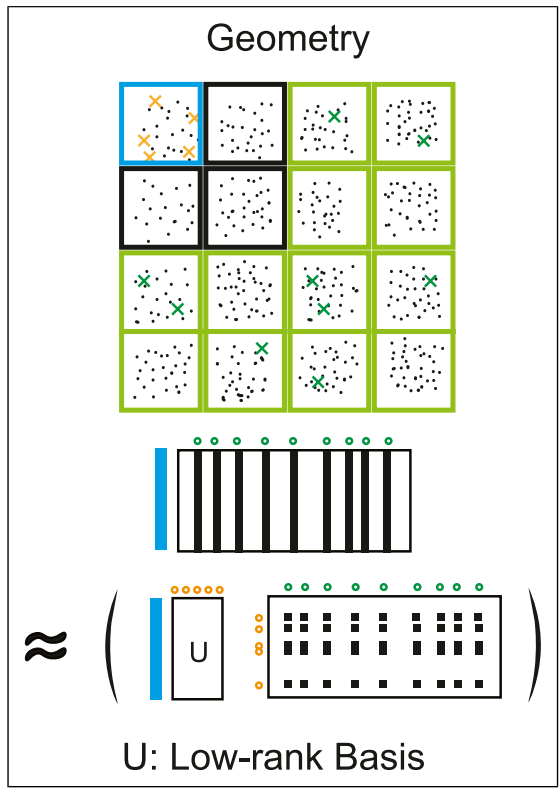

**Figure 7.** Interpolative decomposition on well-separated boxes.

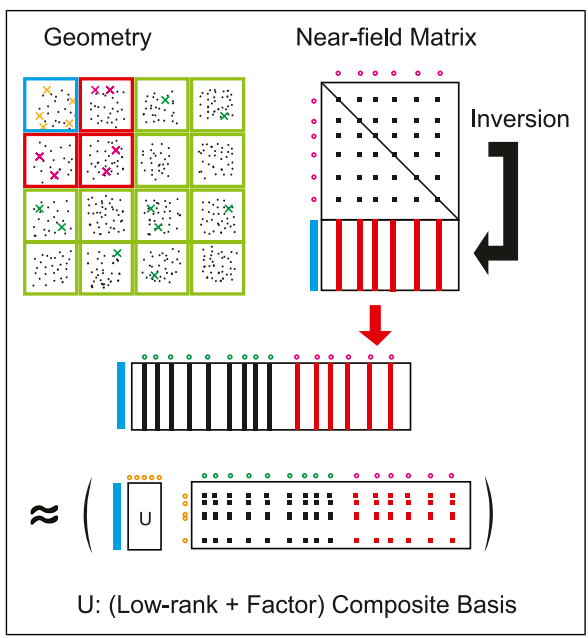

**Figure 8.** Interpolative decomposition on both well-separated and close boxes.

---

**Data**: Geometry info with boxes $B_0, B_1, B_2...$
**Data**: Green's Function $G(x, y)$
**Result**: $\mathcal{H}^2$-matrix $A$ with shared basis $U$
**Result**: Skeletons $SK_i$ for each box $B_i$
**Result**: Coupling matrices $S$

1   **for** $l \leftarrow levels$ **to** 1 **do**
2     **for** *all $U_i$ in level $l$* **do**
3       $S_F \leftarrow$ Sample($\forall B_j$ well-separated with $B_i$);
4       $S_C \leftarrow$ Sample($\forall B_j$ close to $B_i$);
5       $A_{c,c} \leftarrow G(S_C, S_C)$;
6       $A_{Far} \leftarrow G(B_i, S_F)$;
7       $A_{Close} \leftarrow G(B_i, S_C)A_{c,c}^{-1}$;
8       $(U_i, SK_i) \leftarrow$ ID($[A_{Far}, A_{Close}]$);
9     **end**
10    **for** *all $A_{ij}$ in level $l$ where $B_i$ and $B_j$ are close* **do**
11      $A_{ij} \leftarrow G(B_i, B_j)$;
12    **end**
13    **for** *all $S_{ij}$ in level $l$ where $B_i$ and $B_j$ are well-separated* **do**
14      $S_{ij} \leftarrow G(SK_i, SK_j)$;
15    **end**
16    **for** *all $B_i^{l-1}$ in level $l-1$* **do**
17      $B_i^{l-1} \leftarrow \begin{pmatrix} SK_{2i} & SK_{2i+1} \end{pmatrix}$;
18    **end**
19 **end**

**Algorithm 1:** $\mathcal{H}^2$-matrix construction including pre-factorization

---

### 3.6. Parallel $\mathcal{H}^2$-ULV factorization

In this section, we cover the $\mathcal{H}^2$-ULV factorization phase. The factorization algorithm is formulated as a level-wise algorithm, where the factorization within each level is completely parallel. However, it does require synchronized merging and reduction operations between levels. By making the factorization inherently parallel within each level, we expose a large parallel region that can take advantage of batched execution on GPUs. Even when comparing with an implementation of the inherently parallel HSS-ULV factorization, we also expect better parallel scalability due to higher numeric operation intensity in the dense off-diagonal blocks for our $\mathcal{H}^2$-ULV factorization.

**Data**: Symmetric $\mathcal{H}^2$-matrix $A$ with shared basis $U$
**Data**: Coupling matrices $S$

1   **for** $l \leftarrow levels$ **to** $1$ **do**
2    **for** *all $A_{ij}$ in level $l$* **do**
3      $A_{ij}^{SS}, A_{ij}^{RS}, A_{ij}^{SR}, A_{ij}^{RR} \leftarrow U_i^{-1} A_{ij} U_j^{T-1}$;
4    **end**
5    **for** *all $S_{ij}$ in level $l$* **do**
6      $A_{ij}^{SS} \leftarrow S_{ij}$;
7    **end**
8    **for** *all $A_{ii}^{RR}$ in level $l$* **do**
9      $L(r)_{ii}, L(r)_{ii}^T \leftarrow cholesky(A_{ii}^{RR})$;
10     **for** *all $A_{ji}^{RR}$ where $j > i$* **do**
11       $L(r)_{ji} \leftarrow A_{ji}^{RR} L(r)_{ii}^{T-1}$;
12     **end**
13     **for** *all $A_{ji}^{SR}$* **do**
14       $L(s)_{ji} \leftarrow A_{ji}^{SR} L(r)_{ii}^{T-1}$;
15     **end**
16     $A_{ii}^{SS} \leftarrow A_{ii}^{SS} - L(s)_{ii} L(s)_{ii}^T$;
17    **end**
18    **for** *all $A_{ij}^{l-1}$ in level $l-1$* **do**
19     $A_{ij}^{l-1} \leftarrow \begin{pmatrix} A_{2i,2j}^{SS} & A_{2i,2j+1}^{SS} \\ A_{2i+1,2j}^{SS} & A_{2i+1,2j+1}^{SS} \end{pmatrix}$;
20    **end**
21 **end**
22 $L_{00}, L_{00}^T \leftarrow cholesky(A_{00})$;

**Algorithm 2**: $\mathcal{H}^2$-matrix factorization

The details of the factorization phase are shown in Algorithm 2. At this phase we assume that the shared basis that includes both the factorization basis and the low-rank basis has been computed. We start the factorization from the bottom most level and treat each level as a $\mathrm{BLR}^2$ matrix and apply the ULV factorization to it. The ULV factorization of each level consists of a sparsification step and factorization step. In the sparsification step, the transpositions (inverses) of the shared basis are multiplied to the dense matrix to obtain a block-sparse matrix, as shown in Figure 1. In the factorization step, a block-Cholesky factorization is performed on the block-sparse matrix. As we have covered in the sections relating to the factorization basis, the individual sections of $A^{RR}$, $A^{SR}$, $A^{RS}$, and $A^{SS}$ does not require an explicit permutation to align with the block factorization order. We apply the factorization only to the corresponding locations in the partitioned matrix and do the same when applying the forward and backward substitution.

In our parallel ULV factorization, we intentionally skip all Schur complements calculations except for the self-to-self redundant to skeleton Schur complement (on line 28). As we have covered in equation (21), this is the only trailing update, and no other trailing updates are computed nor recompressed. The same algorithmic complexity analysis for other $\mathcal{H}^2$-ULV factorizations also applies to our method, except we can do it in an inherently parallel manner.

## 3.7. Parallel forward and backward substitution

The forward and backward substitution applies the operations $x = (L^T)^{-1} L^{-1} b$ to obtain the solution to the linear system. Substitution is an essential part of the direct solver or preconditioner, but has been difficult to parallelize due to its inherently serial nature. Although the HSS-ULV factorization by Chandrasekaran et al. (2006) and the $\mathcal{H}^2$−$ULV$ factorization by Ma et al. (2022) can process the factorization phase in an inherently parallel manner, these methods do not provide any means to parallelize the substitution phase. In the present work, we show for the first time that the substitution phase can also be parallelized. We only provide the algorithm for performing the forward substitution, since the backward substitution can be applied in a similar manner.

**Data**: ULV-factorized $\mathcal{H}^2$-matrix $A = ULL^T U^T$
**Data**: Vector $b$
**Result**: Vector $y = L^{-1} U^{-1} b$ overwritten in $b$

1   **for** $l \leftarrow levels$ **to** $1$ **do**
2    **for** *all $b_i$ in level $l$* **do**
3     $\begin{pmatrix} b_i^R \\ b_i^S \end{pmatrix} \leftarrow U_i^{-1} b_i$;
4    **end**
5    **for** *all $L_{ii}^{RR}$ in level $l$* **do**
6     $b_i^R \leftarrow (L_{ii}^{RR})^{-1} b_i^R$;
7     **for** *all $L_{ji}^{RR}$ where $j > i$* **do**
8       $b_j^R \leftarrow b_j^R - L_{ji}^{RR} b_i^R$;
9     **end**
10    **for** *all $L_{ji}^{SR}$* **do**
11       $b_j^S \leftarrow b_j^S - L_{ji}^{SR} b_i^R$;
12     **end**
13    **end**
14    **for** *all $b_i^{l-1}$ in level $l-1$* **do**
15     $b_i^{l-1} \leftarrow \begin{pmatrix} b_{2i}^S \\ b_{2i+1}^S \end{pmatrix}$;
16    **end**
17 **end**

**Algorithm 3**: A naïve $\mathcal{H}^2$-matrix forward substitution.

We first describe the naïve method for forward substitution of an $\mathcal{H}^2$-matrix based on the block TRSV algorithm in Algorithm 3. The loop on line number 16 introduces a write after write data dependency and also causes a read after write dependency for the trailing $b_i^R$ in line number 12. When using the block version of the TRSV/TRSM algorithm, the forward and backward substitution will result in an inherently serial algorithm. However, closer examination of the substitution algorithm reveals that, the inverse computation of the triangular factors is a block matrix with low-rank structures, which we can further apply the Schur complements computations to. In other words, we can use a low-rank approximation of $L^{-1}$ and $(L^T)^{-1}$, to avoid the update after update data dependency. In order to parallelize the substitution phase, we take advantage of the following two properties:

- The write after write and read after write dependency affect only the $RR$ regions of the matrix
- The existence of the off-diagonal dense matrix is the source of such data dependencies. (e.g., a diagonal block $RR$ structure in HSS matrices does not have this data dependency)

Because the portions of the vector $b$ corresponding to the skeleton are not being solved until the next level, the updates can happen later than the redundant portion and be processed in parallel. We demonstrate this process using a $4 \times 4$ lower-triangular block matrix with all of the off-diagonal blocks filled with dense matrix blocks that correspond with the $RR$ portions of $LL^T$. This demonstrates the triangular matrix inversion for the most challenging case to parallelize. Even for such a case, we show that our algorithm still has a large degree of parallelism.

$$
L = \begin{pmatrix} L_{00} & & & \\ A_{10}\left(L_{00}^T\right)^{-1} & L_{11} & & \\ A_{20}\left(L_{00}^T\right)^{-1} & A_{21}\left(L_{11}^T\right)^{-1} & L_{22} & \\ A_{30}\left(L_{00}^T\right)^{-1} & A_{31}\left(L_{11}^T\right)^{-1} & A_{32}\left(L_{22}^T\right)^{-1} & L_{33} \end{pmatrix}
\tag{24}
$$

The $L$ factor shown above is constructed with the zeroed redundant trailing fill-ins described in equation (21). The $L_{ii}$ blocks are the actual factors of $A_{ii} = L_{ii}L_{ii}^T$ due to the absence of the trailing updates. To invert the $L$ factor, we decompose the block matrix into a product of lower-triangular matrices, where each multiplicand contains one block column of the original $L$, as $L = L_0L_1L_2L_3$. The inversion of the complete $L$ factor consists of the separate inversion of the four multiplicand matrices and multiplied in reverse order, as $L^{-1} = (L_0L_1L_2L_3)^{-1} = L_3^{-1}L_2^{-1}L_1^{-1}L_0^{-1}$. For simplicity, we first consider only the contents of $L_0$ and $L_1$. The inversion of $L_2$ and $L_3$ factors can be derived following similar rules. The contents of $L_0$ and $L_1$ are shown below:

$$
L_0 = \begin{pmatrix} L_{00} & & & \\ A_{10}\left(L_{00}^T\right)^{-1} & I & & \\ A_{20}\left(L_{00}^T\right)^{-1} & & I & \\ A_{30}\left(L_{00}^T\right)^{-1} & & & I \end{pmatrix}
\tag{25}
$$

$$
L_1 = \begin{pmatrix} I & & & \\ & L_{11} & & \\ & A_{21}\left(L_{11}^T\right)^{-1} & I & \\ & A_{31}\left(L_{11}^T\right)^{-1} & & I \end{pmatrix}
\tag{26}
$$

Inverting the $L_0$ and $L_1$ separately each gives

$$
L_0^{-1} = \begin{pmatrix} L_{00}^{-1} & & & \\ -A_{10}A_{00}^{-1} & I & & \\ -A_{20}A_{00}^{-1} & & I & \\ -A_{30}A_{00}^{-1} & & & I \end{pmatrix}
\tag{27}
$$

$$
L_1^{-1} = \begin{pmatrix} I & & & \\ & L_{11}^{-1} & & \\ & -A_{21}A_{11}^{-1} & I & \\ & -A_{31}A_{11}^{-1} & & I \end{pmatrix}
\tag{28}
$$

In these inverted triangular factors, we merge the product of $\left(L_{ii}^T\right)^{-1}L_{ii}^{-1} = A_{ii}^{-1}$ for $i = 0, 1$. Multiplying $L_1^{-1}L_0^{-1}$ gives

$$
L_1^{-1}L_0^{-1} = \begin{pmatrix} L_{00}^{-1} & & & \\ -L_{11}^{-1}A_{10}A_{00}^{-1} & L_{11}^{-1} & & \\ -A_{20}A_{00}^{-1} + C_{20}A_{00}^{-1} & -A_{21}A_{11}^{-1} & I & \\ -A_{30}A_{00}^{-1} + C_{30}A_{00}^{-1} & -A_{31}A_{11}^{-1} & & I \end{pmatrix}
\tag{29}
$$

where $C_{20} = A_{21}A_{11}^{-1}A_{10}$ and $C_{30} = A_{31}A_{11}^{-1}A_{10}$, which are the Schur complement update term from the elimination of block $A_{11}$. As the matrix notation $A$ corresponds to $RR$ regions of the matrix factor, the intermediate results $C_{20}$ and $C_{30}$ are both zero matrix blocks as can be seen from equation (21). Rewriting the intermediate product of $L_1^{-1}L_0^{-1}$ using the zeroed trailing redundant fill-ins and transforming $L_{ij} = A_{ij}\left(L_{jj}^T\right)^{-1}$ gives

$$
L_1^{-1}L_0^{-1} = \begin{pmatrix} L_{00}^{-1} & & & \\ -L_{11}^{-1}L_{10}L_{00}^{-1} & L_{11}^{-1} & & \\ -L_{20}L_{00}^{-1} & -L_{21}L_{11}^{-1} & I & \\ -L_{30}L_{00}^{-1} & -L_{31}L_{11}^{-1} & & I \end{pmatrix}
\tag{30}
$$

Performing a similar operation on $L_2^{-1}$ and $L_3^{-1}$ gives us the final form of $L^{-1}$:

$$\begin{pmatrix} L_{00}^{-1} & & & \\ -L_{11}^{-1}L_{10}L_{00}^{-1} & L_{11}^{-1} & & \\ -L_{22}^{-1}L_{20}L_{00}^{-1} & -L_{22}^{-1}L_{21}L_{11}^{-1} & L_{22}^{-1} & \\ -L_{33}^{-1}L_{30}L_{00}^{-1} & -L_{33}^{-1}L_{31}L_{11}^{-1} & -L_{33}^{-1}L_{32}L_{22}^{-1} & L_{33}^{-1} \end{pmatrix}$$

(31)

By applying the forward and backward substitution using this reformed $L^{-1}$, the triangular solves are transformed into matrix-vector multiplications, which removes the data dependency during the substitution. Implementation-wise, we have the option to explicitly compute $L^{-1}$ as in equation (9). We have decided not to do this, because of the number of floating point operations to operate on a dense matrix block. The cost of matrix-matrix operations on dense blocks, compared with operating with vectors, is significantly higher, with no apparent benefit in other aspects, such as in communication patterns for distributed memory or implementation difficulties.

## 4. Design considerations for GPUs

Our hierarchical low-rank factorization code has been developed as a higher-level set of algorithms that internally operates on dense matrix structures using BLAS/LAPACK routines. We have covered in the previous section the algorithms for construction, factorization, and substitution. The smallest unit of computation in these algorithms is single matrix blocks, and the operations applied are defined in BLAS and LAPACK. For this reason, rather than building a custom set of GPU/CPU kernels, we decided to use the highly optimized libraries written for GPUs, such as cuBLAS and MAGMA (Tomov et al., 2010a,b; Dongarra et al., 2014), and OneAPI MKL for CPUs. cuBLAS and cuSOLVER are proprietary libraries that are included with each release of CUDA. They are capable of delivering the best performance on a wide range of basic matrix operations. MAGMA, on the other hand, provides a wider range of interfaces and algorithms, such as batched data copying and batching of variable-sized matrices. OneAPI MKL is part of Intel's implementation of SYCL standard and DPC++ and has extended the original MKL, an x86-64 BLAS and LAPACK implementation, to heterogeneous computing platforms that run with OpenCL. The programming model of SYCL and OneAPI DPC++ promises a bright future for performance portability across platforms and different types of hardware. Fugaku at RIKEN which has A64FX CPUs, Frontier at ORNL which has AMD GPUs, and Aurora at ANL has Intel GPUs. These are all good candidates for cross-platform tools. As our primary interest lies in delivering the best performance with the hardware we currently have, we view the usage of such performance portable programming models as a future step to take. Experiences and optimizations written for CUDA and NVIDIA libraries are also helpful when switching to more complex tools like SYCL for deciding different design considerations, such as choosing from the internal data storage layout or padding and alignment options. In this section, we will discuss the various design decisions we made for the highly optimized implementation of our algorithm on GPUs.

### 4.1. Variable-size batch versus constant-size batch

We have introduced the ULV factorization as an inherently parallel algorithm, which can be implemented using a series of inherently parallel loops. To extract the best performance, these loops need to be executed as batched calls due to the relatively small size of the matrix blocks. The NVIDIA libraries cuBLAS and cuSOLVER provide batched interfaces for most functions that we use, such as Cholesky factorization (POTRF), triangular matrix solve (TRSM), and matrix-matrix multiplications (GEMM). Although delivering very high performance even for relatively small matrix sizes, these batched functions require the batch of matrices to have matching dimensions and strides. On the other hand, MAGMA and the group call in OneAPI MKL provide both constant-size batch interfaces and variably-sized batch interfaces. In our case, batched calls for varying matrix sizes enable us to use adaptive matrix ranks when constructing the $\mathcal{H}^2$-matrix, which results in a significant improvement in both performance and memory efficiency.

We have run benchmarks on these different libraries on our computational workload. Shortly summarizing the performance, we discovered a significant performance degradation when using variable-sized matrix dimensions and strides. Even if running on the same dimension but with a variable-size batching interface, the performance can be roughly 50% slower than the constant-sized version. For this reason, we decided that for most use cases where memory is sufficient, zero-padding the matrix with lower ranks to the maximum in the batch gives the best performance instead of using the variable-sized batch provided by MAGMA, although some computational FLOPs are being wasted from padding.

For processing matrices with different sizes, padding zeros to the unmatched dimensions is a useful technique to take advantage of the batched routines. We align the dimensions of the redundant and the skeleton to the maximum of the level, as well as making this dimension multiples of 4, as the cuBLAS and cuSOLVER libraries have suggested. In our case however, the primary challenge lies in simply padding zeros to the deficient dimensions can only work with the matrix-matrix multiply operations that exist in matrix sparsification as well as Schur complements computations. Completely zeroed-out rows and columns halt the Cholesky factorization computations as it encounters diagonal zero entries and produces division by zero errors when later applied in the triangular matrix solves. During the factorization phase, we used a batched AXPY call that writes entries to the diagonal fill up the zero rows and columns that will appear in the factorization. The cuBLAS and cuSOLVER libraries do not provide a native AXPY interface, so we used the batched GEMM interface instead by setting the input M and K to 1 so that LDB and LDC correspond to the functions of INCX and INCY parameters accordingly.

### 4.2. Optimizations for ULV Factorization

In the previous section, we introduced a ULV factorization based on an internal block Cholesky factorization in Algorithm 2. The naïve implementation already exposed large regions of parallelism, where each diagonal block could be factorized independently. Compared to simple multi-threading with "parallel for," the batched API requires additional considerations as we will describe in the following subsection.

### 4.3. Temporary storage for matrix sparsification

The primary concern for an effective GPU implementation is the amount of temporary storage that is needed for the factorization, which must not exceed the total amount of the total available on board memory. The composite basis $U$ needs to be stored explicitly in full dense matrix format, for accessing the skeleton and the redundant parts of the basis. The matrix sparsification process requires the same amount of temporary storage to store the intermediate result of basis transformation as the storage required by the original form of the matrix. In the ideal case where we can always allocate an abundant amount of temporary storage, we can launch a single batched GEMM call for all basis transformations. A more practical approach that is adopted by our implementation is to allocate the amount needed for the diagonals. We believe that having the amount of temporary storage matching the number of diagonal blocks is a good call for two reasons. Firstly, the diagonal blocks are factorized in parallel and can happen concurrently. If we want to launch a single batched POTRF call for all diagonal blocks, the sparsification process needs to have the $U$ and $V$ transformations of all of the diagonal blocks in the GPU memory. Secondly, the total number of blocks including the off-diagonals for each level is linearly dependent on the number of diagonal blocks. Therefore, even if we cannot launch a single batched GEMM call for all blocks, the number of launched batched GEMMs is bounded by a constant, which is the average number of blocks per column. We give a more detailed and quantitative analysis of this optimization, in the "Numerical Results" section with NVIDIA profiler results.

### 4.4. Reusing intermediate results of TRSM

For entries in the lower triangular factor, two transformations are being applied from the right side to the original entry.

$$\begin{cases} L(r)_{ij} = \left(U_i^R\right)^T A_{ij} U_j^R \left(L(r)_{jj}^t\right)^{-1}, i > j \\ L(s)_{ij} = \left(U_i^S\right)^T A_{ij} U_j^R \left(L(r)_{jj}^t\right)^{-1} \end{cases} \tag{32}$$

For the blocks on the same columns of the lower triangular factor, the term $U_j^R (L(r)_{jj}^t)^{-1}$ can be reused to reduce the number of TRSM operations. The factorization algorithm is described in Algorithm 4, featuring a new temporary storage $V$ to store the intermediate result of $U_j^R (L(r)_{jj}^t)^{-1}$.

---

**Data:** symmetric $\mathcal{H}^2$-matrix $A$ with shared basis $U$
**Data:** coupling matrices $S$

1   **for** $l \leftarrow levels$ **to** 1 **do**
2    **for** *all $A_{ii}$ in level l* **do**
3     $A_{ii}^{SS}, A_{ii}^{RS}, A_{ii}^{SR}, A_{ii}^{RR} \leftarrow U_i^{-1} A_{ii} U_i^{T-1}$;
4     $L(r)_{ii}, L(r)_{ii}^T \leftarrow cholesky(A_{ii}^{RR})$;
5     $L(s)_{ii} \leftarrow A_{ii}^{SR} L(r)_{ii}^{T-1}$;
6     $A_{ii}^{SS} \leftarrow A_{ii}^{SS} - L(s)_{ii} L(s)_{ii}^T$;
7     $V_i \leftarrow U_i^{T-1} L(r)_{ii}^{-1}$;
8    **end**
9    **for** *all $A_{ij}$ in level l where $i \neq j$* **do**
10     $A_{ij}^{SS}, A_{ij}^{RS}, A_{ij}^{SR}, A_{ij}^{RR} \leftarrow U_i^{-1} A_{ij} V_j$;
11     **if** $i > j$ **then**
12      $L(r)_{ij} \leftarrow A_{ij}^{RR}$;
13     **end**
14     $L(s)_{ij} \leftarrow A_{ij}^{SR}$;
15    **end**
16    **for** *all $S_{ij}$ in level l* **do**
17     $A_{ij}^{SS} \leftarrow S_{ij}$;
18    **end**
19    **for** *all $A_{ij}^{l-1}$ in level l − 1* **do**
20     $A_{ij}^{l-1} \leftarrow \begin{pmatrix} A_{2i,2j}^{SS} & A_{2i,2j+1}^{SS} \\ A_{2i+1,2j}^{SS} & A_{2i+1,2j+1}^{SS} \end{pmatrix}$;
21    **end**
22 **end**
23 $L_{00}, L_{00}^T \leftarrow cholesky(A_{00})$;

**Algorithm 4:** Improved $\mathcal{H}^2$-matrix factorization

---

## 5. Distributed-memory implementation

Our algorithm for parallel $\mathcal{H}^2$-ULV factorization and substitution features decoupled diagonal blocks, which allows us to use the simple 1-D row/column partitioning that has been used for fast multipole method (FMM) and HSS implementations for distributed memory. This is different from distributed memory implementations of dense direct solvers like ScaLAPACK, which use block cyclic partitioning. The primary reason for using block cyclic partitioning is to maximize load balance when there are dependencies on the trailing updates. Our $\mathcal{H}^2$-ULV factorization removes such dependencies, so there is no need to use a block cyclic partitioning. Using the simple row/column partitioning scheme has dramatically simplified our implementation, and allows us to use the same structure as a shared memory implementation for most of our distributed implementation.

For the libraries chosen, limited support for MPI and direct means for device-to-device communications in the SYCL standards and Intel's implementation of DPC++ has also been a factor for us to consider in the long term. For native CUDA applications, although we are fully aware of the existence of CUDA-aware MPI implementations such as OpenMPI and MVAPICH, we choose to use NCCL for communicating among GPUs. The primary reason for using NCCL over MPI is the better integration with the CUDA programming model of less host-device synchronization overhead before initiating any collective communications. NCCL communicators are fully integrated with cuda-Streams and launch explicit kernels for communication and peer access, which automatically resolves the compute-communication data dependency within the stream. One drawback of using NCCL is that it does not have all of the collectives that the MPI standard defines. Another issue is that it limits the portability to other distributed memory platforms that do not have NVIDIA GPUs. NCCL also does not support AllGatherv, but our choice to use a constant rank during the batched matrix operations allows us to get away with AllGather.

In order to minimize communication, it is necessary to map the proximity in the underlying geometry to the proximity of the process distribution. Though for higher dimensional problems, such a mapping is not possible, and there will always be a set of points that are close in the geometry but placed far apart in the process distribution. The use of space-filling curves has dramatically reduced the number of neighbor communications compared to using any cyclic process distributions. In our implementation, we used a 1-D column process distribution for better data-locality in the $LL^T$ factorization, where the diagonal $(L^T)^{-1}$ needs to be applied to all of the off-diagonal blocks located on the same column. If the internal operations are factored into the upper triangular form $U_*^T U_*$, using the 1-D row process distribution will become more advantageous, as they share a significant commonality.

### 5.1. Communications in factorization

In the $\mathcal{H}^2$-ULV factorization, the upper levels have relatively small amount of computation, but could result in a large amount of communication if not implemented carefully. We use a common technique for hierarchical algorithms such as FMM and multigrid, where a hierarchical AllGather is used to merge the contents while computing the upper levels redundantly. The hierarchical AllGather forms a binary tree with split communicators at each level, as described in Figure 9 along with the process distribution pattern. Although some contents in each node might be duplicated, the tree representation of the $\mathcal{H}^2$-matrix representation remains distributed across all available processing units. More precisely, the AllGather is required when performing the following operation:

$$A_{ij}^{l-1} : P_{0,1} \leftarrow \begin{pmatrix} A_{2i,2j}^{SS} : P_0 & A_{2i,2j+1}^{SS} : P_1 \\ A_{2i+1,2j}^{SS} : P_0 & A_{2i+1,2j+1}^{SS} : P_1 \end{pmatrix} \qquad (33)$$

which is the last step of the $\mathcal{H}^2$-ULV factorization algorithm inside each level. Column $2j$ and column $2j + 1$ are computed and stored on different processor ranks. To minimize the data movement, we directly perform memory writes, and also store the low-rank coupling matrices directly to the merged locations. An alternative to the AllGather collective is to use AllReduce to the merged block on the next level. Using AllReduce greatly simplifies the implementation as shown in equation (34):

$$A_{ij}^{l-1} \leftarrow \begin{pmatrix} A_{2i,2j}^{SS} & 0 \\ A_{2i+1,2j}^{SS} & 0 \end{pmatrix} : P_0 + \begin{pmatrix} 0 & A_{2i,2j+1}^{SS} \\ 0 & A_{2i+1,2j+1}^{SS} \end{pmatrix} : P_1 \quad (34)$$

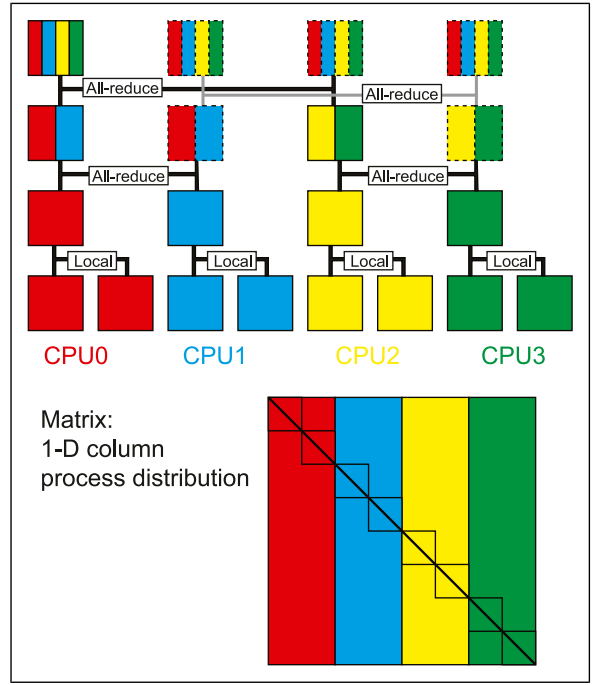

**Figure 9.** Merging pattern for leftover $A^{SS}$ blocks from ULV-factorization.

Two subjects in this approach for distributed-memory communication are worth noting that show significance to the results. Firstly, for factorization only, both the number of collective communication function calls and the message sizes are independent of the problem size $N$. This is true because the matrix factorization for levels under $\log_2 P$ levels below the root is computed completely locally inside each processing unit. In a more extreme comparison, the factorization of a $\mathcal{H}^2$-matrix with two levels using four processing units, has the same communication complexity as factorizing another $\mathcal{H}^2$-matrix with 20 or even 200 levels with the same leaf size and admissibility condition, despite having 10x and 100x difference in memory and algorithmic complexities.

Secondly, the processes holding the gathered or reduced content perform redundant computation on the replicated data. The amount of replicated data and redundant computation is dependent on the number of levels, as well as the number of available processes. The total amount of work computed beyond level $l = \log_2 P$ is $\mathcal{O}(P \log_2 P)$, where $P$ is the number of processors. Among this work, only $\mathcal{O}(2^{\log_2 P}) = \mathcal{O}(P)$ is distinctive. Although for a fixed number of computing units, this number becomes a constant, with an increasing number of processes, this factor of $\mathcal{O}(P \log_2 P)$ becomes significant.

The redundant computation and replicated data during the factorization become useful during the forward and backward substitution. Holding the same content in different processes saves the effort of broadcasting the split partitions to the children when moving from the root to the leaf during the substitution. The gradual reduction in parallel regions for levels beyond level $= \log_2 P$ also makes many processors idle if the redundant work is not performed. In other words, the redundant computation utilizes the otherwise idle compute resources in order to reduce communication.

## 5.2. Communications in substitution

Forward substitution features the same merging strategy as that of the factorization. The substitution algorithm also requires neighbor-to-neighbor communications, which is similar to that of FMM and $\mathcal{H}^2$ matrix-vector multiplication. Compared to the factorization, which only performs merge operations between levels and has very high intensity of floating point operations, the substitution is significantly more memory and communication bound even for our inherently parallel version. In addition to the previously used merging communication when moving between levels of the $\mathcal{H}^2$-matrix, two different patterns of communication are present during the substitution; (1) summing the updated contents among neighbors, and (2) broadcasting the finalized results to neighbors, as shown in Figure 10. Despite the difference in the type of collective, we can reuse the same communicator for these collectives. In fact, the same neighbor communicator can be reused even for $\mathcal{H}^2$ matrix-vector multiplication as well as in $\mathcal{H}^2$-matrix construction.

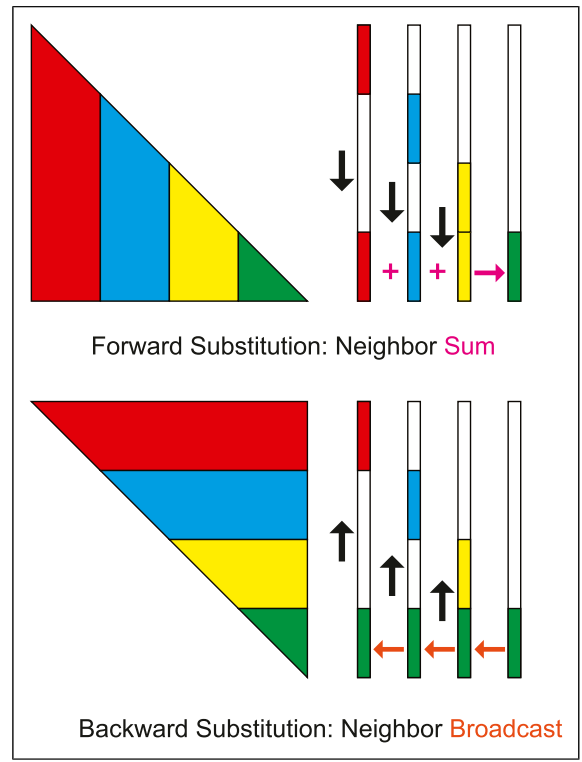

**Figure 10.** Neighbor communication pattern for forward and backward substitution.

As the number of closely interacting neighbors in a geometry configuration with a fixed admissibility condition is constant, the cost of neighbor communications can be conditionally bounded by a constant. Both the summation (reduction) and broadcast message sizes are dependent on the size of the locally computed content. The size of the substituted local vector has a length of $N/P$, being the same as the overall memory complexity $\mathcal{O}(N/P)$ of the $\mathcal{H}^2$-matrix. In the weak scaling experiments where the amount of work distributed to each processing unit is the same, $\mathcal{O}(N/P)$ becomes $\mathcal{O}(1)$. As a result of communicating on all levels and interleaving with the substitution computations, the neighboring communications appear to be more costly latency-wise compared to the simple reduce-broadcast communication pattern when merging between levels. This is also a contributing factor to the eventual loss of scalability in the forward and backward substitution algorithms when compared with the factorization.

In the previous section, we claimed that explicitly forming $L^{-1}$ does not simplify the communications, as the neighboring communication is still needed due to the diagonal contents not being present in all processes for computing the explicit inversion of the off-diagonal blocks in equation (31). Our approach to the forward and backward substitution showed a significant impact on the neighboring communication patterns. For the naïve block-TRSV-based algorithm, both the reduction and the broadcasting of the solved results are serialized due to the data dependencies from the previous blocks. Our parallel approach uses only TRSV on the diagonal blocks, and the off-diagonal computations have been modified into a triangular matrix-vector multiplication. This transition from a triangular solve to a matrix-vector multiplication removes the data dependency during the substitution. This greatly reduces the idling of processes and also enables separate communication kernel launching to achieve even lower latency.

## 6. Numerical results

In the section, we present the numerical results on two different test cases; (1) 3-D Laplace equation on a uniformly distributed spherical surface geometry and (2) 3-D Yukawa potential on Hemoglobin molecule (Figure 11). As there have been very few attempts to run any hierarchical low-rank matrix factorization on GPUs in distributed memory environments, the only competitive libary we could find was LORAPO—a library that performs Cholesky factorization on block low-rank (BLR) matrices. Our code was implemented in C++ and CUDA, with only dependencies to dense linear algebra and MPI libraries. Experiments are carried out on the ABCI system at AIST in Japan, and both our code and LORAPO have been compiled with GCC 9.3.0, CUDA 11.3, Open-MPI 4.1.3, Intel MKL 2022, and NCCL 2.9.

We organize the following sections into two parts: (1) the verification of the computational complexity and numerical accuracy and (2) the parallel scalability. For the larger scale experiments, we used up to 128 of the compute V nodes on ABCI, featuring 2x Intel Xeon Gold 6148 Processor running at 2.4 GHz, 20 Cores (40 Threads), and 4x NVIDIA V100 16GiB HBM2 interconnected with NVLink within each node. The single-node experiments were performed on privately owned NVIDIA A100 GPUs and AMD EPYC 7502 CPUs.

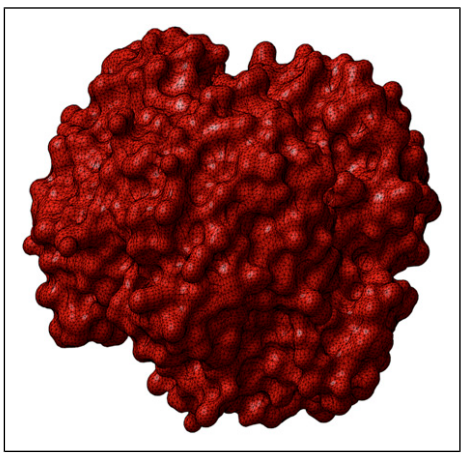

**Figure 11.** A single hemoglobin molecule.

### 6.1. Profiling results

Before analyzing the numerical results and actual timings of the experiments, we present the output from a profiler, which shows the behavior of our proposed algorithm and what performance to expect. We used the profiler tools provided by NVIDIA to check the behavior as well as GPU utilization. Figure 12 shows an annotated screenshot of only the factorization part running on four NVIDIA A100 GPUs for a matrix size of $N = 262, 144$. The profiler results show that the four GPUs show high concurrency and utilization. The GPU utilization starts to fluctuate at higher levels of the tree, but remains high throughout the entire execution. The batched calls to POTRF, TRSM, and GEMM are shown using a colored bar only for GPU1, but the other GPUs are also calling the same libraries at similar timings.

### 6.2. Single node performance and algorithmic complexity

The computation time for the factorization and substitution phase of our $\mathcal{H}^2$-ULV method is shown in Figure 13. The blue lines indicate the results on an AMD EPYC 7502 CPU and the red lines indicate the results on an NVIDIA A100 GPU. For these experiments, we used a dense matrix generated from a 3-D Laplace equation with points distributed on a spherical surface. This setup places the mesh points evenly on the spherical surface with roughly equal spacing for simplicity. The discrete solution to the 3-D Laplace equation is obtained via the Green's function of the Laplace operator, and the individual matrix entry can be obtained through the following kernel, where $r_{ij}$ stands for the 3-D Euclidean distance between the $i$th and $j$th particle in the sorted list:

$$A_{ij} = \begin{cases} 1.E^3, \text{if } i = j \\ \dfrac{1.}{r_{ij}}, \text{if } i \neq j \end{cases} \quad (35)$$

Our method shows the expected $\mathcal{O}(N)$ algorithmic complexity with respect to the matrix dimension $N$. For the CPU, we utilize the multiple cores by launching as many processes as there are cores, while for the GPU we launch a single process and call the batched libraries in cuBLAS/cuSOLVER. The substitution phase is considerably less compute-intensive and is memory-bound. Further, the substitution on the CPU does not use our inherently parallel algorithm, so the computational time is close to that of the factorization despite the significantly smaller workload. The substitution on the GPU uses our inherently parallel algorithm and utilizes the batched kernels. The substitution time scales sub-linearly with respect to the increasing problem size, which shows that the algorithm is capable of utilizing more and more computing capabilities from the GPU as the parallel workload intensifies. For a large enough matrix size, the substitution on the GPU is an order of magnitude faster than the factorization. This would not have been possible without our inherently parallel substitution algorithm.

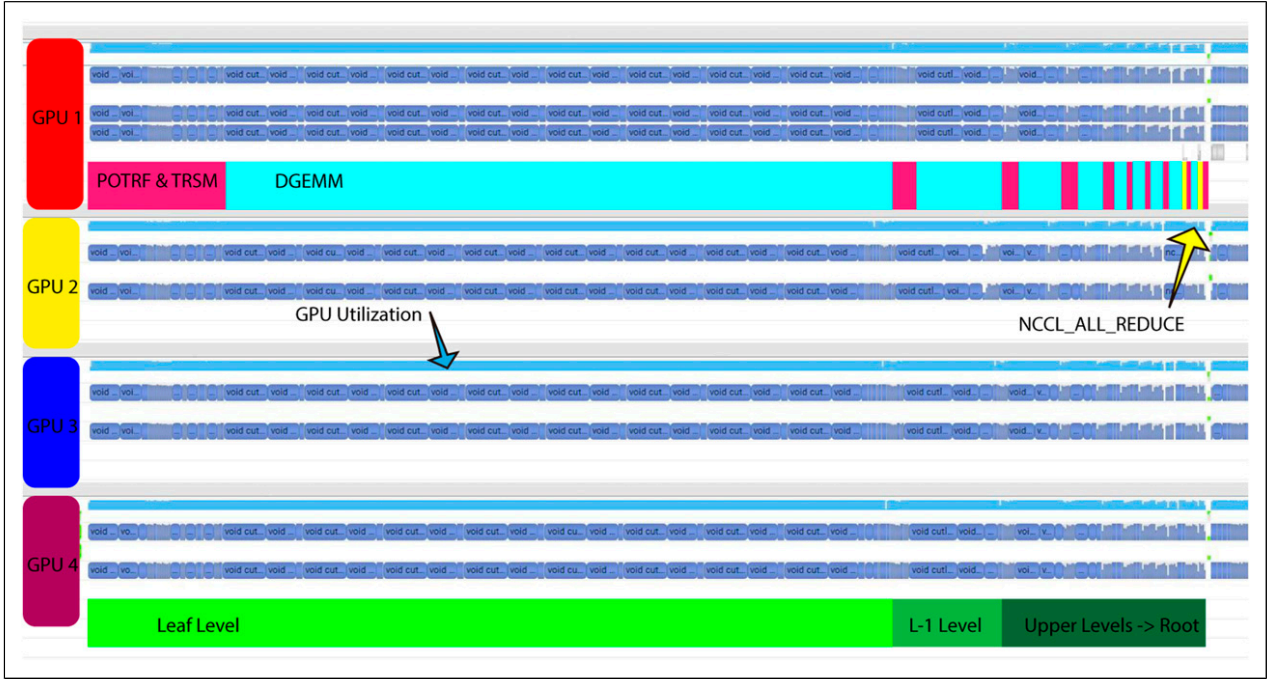

**Figure 12.** Nsys profile screenshot for factorization on 4× A100.

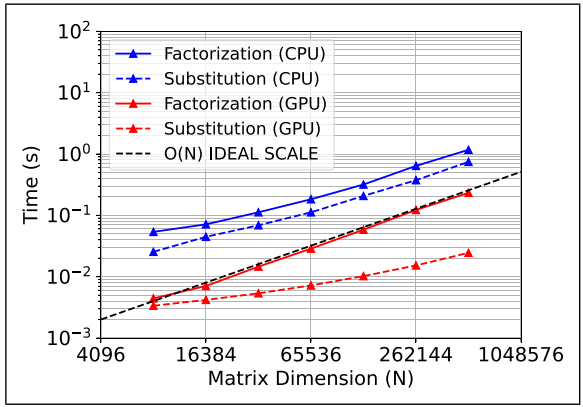

**Figure 13:** $\mathcal{O}(N)$ factorization time.

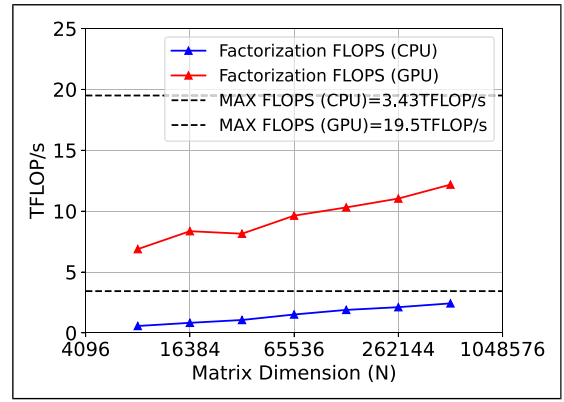

**Figure 14.** Matrix dimension versus Factorization FLOPS performance.

As a more precise indicator of the hardware utilization, we show the arithmetic throughput [TFLOP/s] for the factorization phase in Figure 14. The highest achieved performance from the single node factorization experiments shows a maximum of 2.42TFLOP/s on CPU and 12.18TFLOP/s on GPU. These numbers correspond to 70.6% and 62.5% of the maximum achievable performance on this hardware, which is 3.43 TFLOP/s (on AVX2) and 19.5 TFLOP/s (on FP64 Tensor Cores), respectively.

As a supplement to the arithmetic complexity, we also counted the number of floating point operations that were performed in these experiments in Figure 15. We plotted both the ideal $\mathcal{O}(N)$ and $\mathcal{O}(N \log_2 N)$ reference lines in the figure. The slope of our $\mathcal{H}^2$-ULV factorization seems to be somewhere in between $\mathcal{O}(N)$ and $\mathcal{O}(N \log_2 N)$, and not strictly $\mathcal{O}(N)$. The timing results in Figure 13 obfuscate this fact, due to the under-utilization of hardware for smaller matrix sizes. In order to investigate this deviation from the ideal $\mathcal{O}(N)$ complexity, we looked at the number of dense matrix blocks $N_{NZB}$ as a function of the number of blocks at the leaf level. The number of dense matrix blocks is equal to the number of neighbor interactions, which is what we show in Figure 16. When both leaf size and the admissibility condition are fixed, $N_{NZB} = \mathcal{O}(N)$ that scales linearly with respect to the matrix size, and has a theoretical upper bound. When the number of leaf-level boxes is small, or in other words, the level of the tree is shallow, the actual number of neighboring interactions is much lower than the theoretical upper bound. This is the reason why the number of arithmetic operations in Figure 15 shows a trend that is closer to $\mathcal{O}(N \log_2 N)$ instead of $\mathcal{O}(N)$. By extending the problem size and levels of the partitioning, we can observe that the complexity becomes closer to $\mathcal{O}(N)$ eventually.

In the last figure of this section (Figure 17), we compared the cost of the two separate factorization steps. In the experiments, we used the matrix size where $N = 262,144$ and adjusted the admissibility condition number from 0.0 (HSS admissibility) to 3.0 to produce $\mathcal{H}^2$-matrices with different numbers of off-diagonal dense blocks. The admissibility condition number is defined as the acceptable ratio of the maximum radius and the center distances of the two different boxes of particles. The comparison is made in the number of FP64 operations instead of providing the exact timings of either the pre-factorization or the $\mathcal{H}^2$-matrix construction. The pre-factorization does not require a specific selection of the low-rank approximation method, and such difference in $\mathcal{H}^2$-matrix approximation can create a significant difference in the achievable accuracy and performance. It is also worth noting that the factorization costs, both the pre-factorization and the actual factorization, do not scale linearly with respect to the admissibility condition. In other words, the linear complexity experiments need to keep the admissibility condition number constant. However, this factorization cost comparison result suggests that as the $\mathcal{H}^2$-matrices can have more off-diagonal dense blocks, the pre-factorization does not cost more than 46% of the total cost, as well as scaling linearly just as the actual factorization.

### 6.3. Matrix rank and solution accuracy

There are existing parallel implementations of the HSS-ULV factorization, such as STRUMPACK and H2pack. As a justification for extending this to $\mathcal{H}^2$-matrix format with stronger admissibility conditions at the cost of increased code complexity, we compare the solution accuracy from the different admissibility conditions, as well as the compute time and the number of floating point operations being used to obtain the answers. In these experiments, we used the same code but different runtime parameters for the admissibility condition for computing in the HSS and $\mathcal{H}^2$ matrix formats. The HSS matrix is a subset of the more general $\mathcal{H}^2$ matrix, and can be configured as an $\mathcal{H}^2$-matrix where all of the off-diagonal blocks are low-rank matrices. Although we are fully aware of other existing HSS-matrix libraries that can run on GPUs, we used our implementation for this comparison to avoid contaminating the results with other differences such as the method of low-rank approximation and the level of optimization.

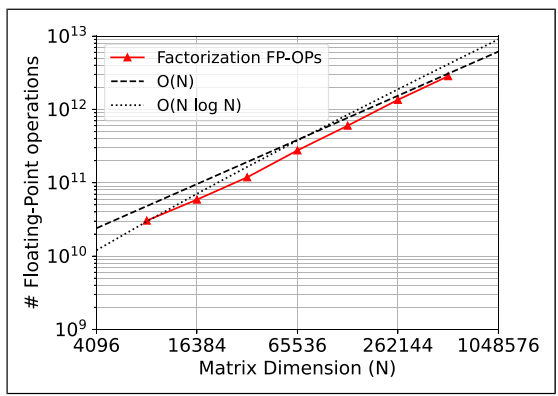

**Figure 15.** Matrix dimension versus Factorization FLOPS count.

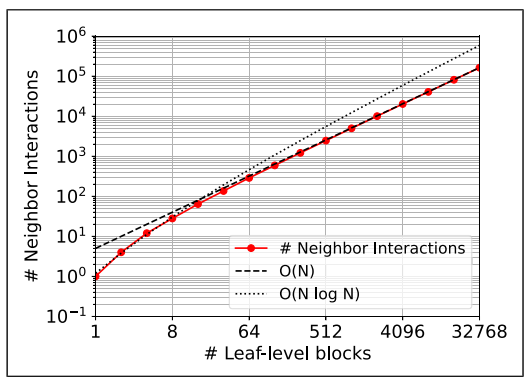

**Figure 16.** Number of neighboring interactions.

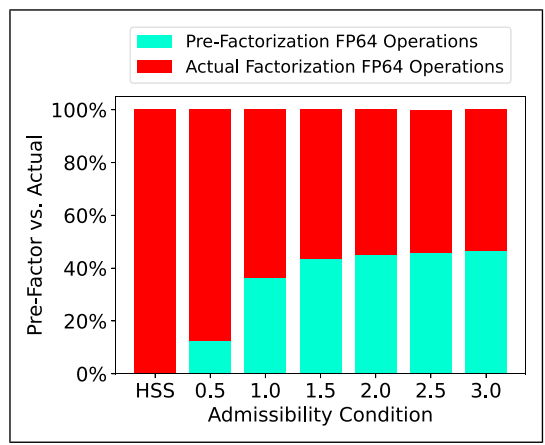

**Figure 17.** Factorization FLOPS distribution versus admissibility condition.

We used a matrix of size $N = 8192$ and *Leaf* = 512 and configured the matrix construction to use a fixed-rank truncation on all of the available far-field particles (far-field sampling disabled, leading to $\mathcal{O}(N^2)$ cost to construct), to give the best low-rank approximation results. Figure 18 shows the result from our experiments, which suggests that in order to achieve the same solution accuracy of that of the $\mathcal{H}^2$-matrix at rank 50, the HSS matrix will require a rank of more than 400. Despite that our method of constructing the $\mathcal{H}^2$-matrix includes both the content from the far-field (low-rank basis) and the neighboring boxes (factorization basis), our method gives considerably better approximation results with small matrix ranks. As the HSS uses very high ranks to achieve the same factorization accuracy, it quickly exhausts the available memory and also results in much slower time-to-solution as can be seen from Figure 19.

## 6.4. Parallel scalability

We now show the parallel scalability results of our $\mathcal{H}^2$-ULV factorization on up to 512 CPU/GPUs on the ABCI system with a maximum problem size of $N = 29,242,368$. In this experiment setup, we compute the Yukawa potential for the boundary mesh points placed on the surface of hemoglobin molecules. The Yukawa potential is similar to Green's function of the Laplace equation, in that the potential at $r = 0$ is a singular point of infinite potential. A simplified Yukawa potential can be computed from the following kernel function by setting all of the constants to 1:

$$A_{ij} = \begin{cases} 1.E^3, \text{if } i = j \\ \dfrac{e^{-r_{ij}}}{r_{ij}}, \text{if } i \neq j \end{cases} \tag{36}$$

The molecule geometries data contains 14,908 and 57,114 mesh points for two differently shaped hemoglobin molecules, and at most 512 duplicates of the same molecule are placed in the same domain. By reading the portions of the geometry of the molecules, we create variations in the problem sizes.

The strong scaling result is presented in Figure 20, where we compare with a BLR-Cholesky factorization implementation in LORAPO. Since the BLR matrix format has a factorization complexity of $\mathcal{O}(N^2)$, we are not able to run LORAPO up to the same problem sizes that we are capable of solving with our method. Our code is executed on the GPU while LORAPO is executed on the CPU. Our code exhibits good strong scaling up to 64 GPUs, but struggled to get close to the ideal scaling past 128 GPUs. This is because the local problem size becomes too small to saturate the performance of the GPUs in such strong scaling experiments. The best performance in the strong scaling experiments showed a speedup of over 13,300x in factorization time over LORAPO using the same number of 128 sockets of CPUs/GPUs.

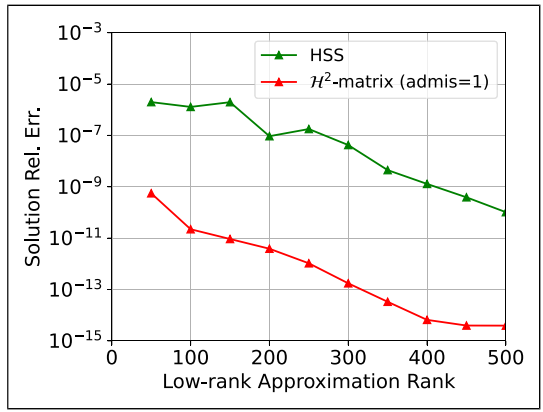

**Figure 18.** Low-rank approximation rank versus solution accuracy.

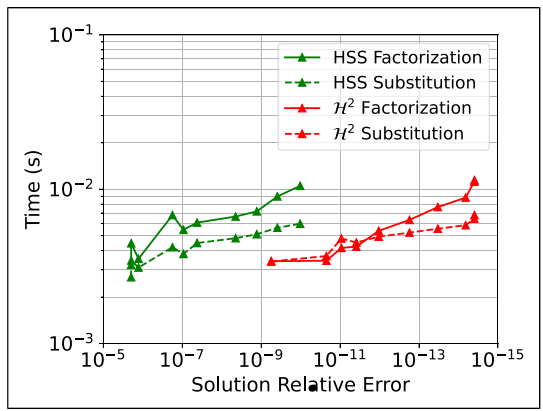

**Figure 19.** Solution accuracy versus time to solution.

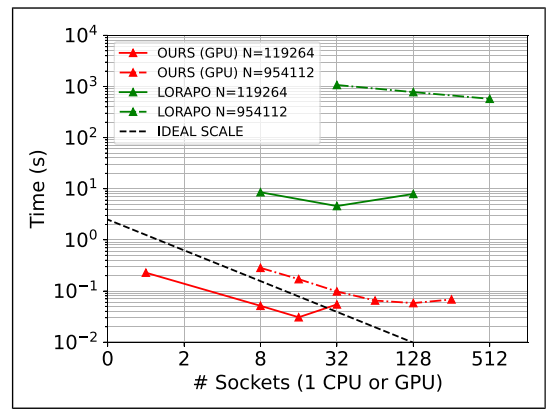

**Figure 20.** Factorization time versus number of CPU/GPUs (strong scaling).

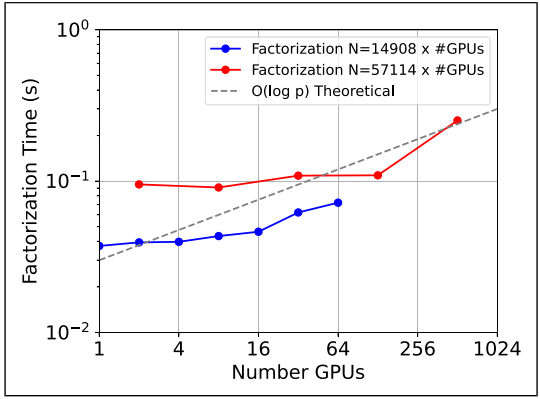

**Figure 21.** Factorization time with increasing computing resource and problem size (weak scaling).

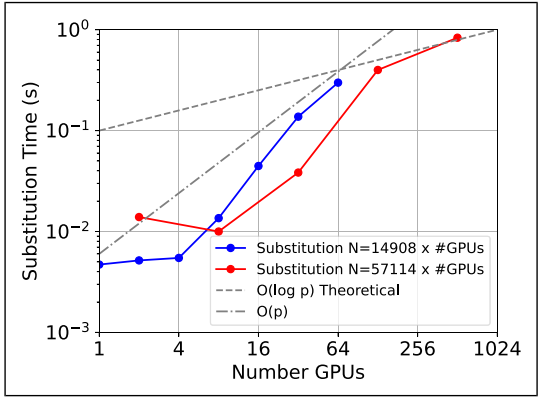

**Figure 22.** Substitution time with increasing computing resource and problem size (weak scaling).

We also further conducted weak scaling experiments to demonstrate our GPU implementation's parallel scalability. Instead of having an ideal constant computing time, our code is expected to have a $\mathcal{O}(\log_2 P)$ compute complexity with a varying problem size where $N = \mathcal{O}(P)$, and $P$ stands for the number of computing processes (resources). This $\mathcal{O}(\log_2 P)$ complexity results from redundant computation at the levels that are close to the root. From levels $\log_2 P$ to level 1, each computing level requires $\mathcal{O}(1)$ work. We present the weak scaling result for the factorization phase in Figure 21, and for the substitution phase in Figure 22. For a maximum problem size of $N = 29, 242, 368$ and using 512 GPUs, we can complete the factorization using only 0.25s and the substitution using 0.83s, utilizing 0.808 PFLOP/s (1.579TFLOP/s per GPU) for factorization. The factorization has high arithmetic intensity and does not need communication until levels that are close to the root. Therefore, it shows almost ideal scaling with respect to the theoretical complexity of $\mathcal{O}(\log_2 P)$. The substitution is memory and communication bound and requires a significant amount of neighbor-to-neighbor communication even at the leaf level. In Figure 22, we plotted both the $\mathcal{O}(\log_2 P)$ and $\mathcal{O}(P)$ scaling lines for reference. As the neighbor-to-neighbor communication intensifies between the nodes, the substitution time grows as $\mathcal{O}(P)$, which is expected. By using a large number of GPUs, the substitution time shows the ideal scaling of $\mathcal{O}(\log_2 P)$.

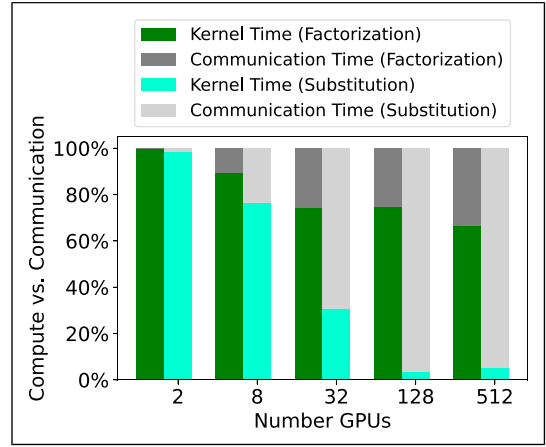

**Figure 23.** Timings percentage breakdown for compute versus communication in weak scaling.

In the weak scaling experiments, the detailed timing breakdown is provided in Figure 23 for experiments ranging from $N = 114, 228$ to $N = 29, 242, 368$. Throughout the different experiment configurations, the factorization kernel time maintained high occupancy (more than 60%) due to its high numeric intensity. For the forward and backward substitution, due to operating on a *width* = 1 vector, the total numbers of FP64 operations are significantly lower compared to the factorization, and the communication costs become dominating as the number of processing units increases. However, since the communication cost has the same complexity as $\mathcal{O}(\log_2 P)$, which is the same as both the factorization and the substitution cost, we can observe that at the eventual point where using 512 GPUs the substitution kernel time ratio does not drop lower comparing to the previous data point of 128 GPUs. From this we conclude, that despite having a high cost for communication, the proposed method maintained its scalability and is expected to scale beyond the parallel resources available to us.

## 7. Conclusion

In this work, we have developed a highly scalable algorithm for an $\mathcal{O}(N)$ Cholesky factorization for rank-structured dense matrices and its multi-GPU implementation. By pre-computing the fill-ins and including them in the shared basis, we are able to devise an inherently parallel factorization algorithm even for strongly admissible $\mathcal{H}^2$-matrices. We also develop a novel algorithm for performing the forward and backward substitution in an inherently parallel manner. This inherently parallel factorization and substitution algorithms allows us to use batched kernels in cuBLAS and cuSOLVER, and extract the full potential GPUs even for small block sizes.

Our method shows superiority to simpler low-rank matrix formats such as BLR and HSS in arithmetic complexity, and the ability to handle higher dimension geometry. To the best of our knowledge, our method is the only inherently parallel algorithm for factorization and substitution for strongly admissible hierarchical low-rank matrices. From the removed dependencies in the factorization trailing panel updates and the substitution vector updates, we are able to solve a 3-D Yukawa potential problem of $N =$ 29,242,368 under 1 s for both factorization and substitution, using 512 NVIDIA V100 GPUs.

## Declaration of conflicting interests

The author(s) declared no potential conflicts of interest with respect to the research, authorship, and/or publication of this article.

## Funding

The author(s) disclosed receipt of the following financial support for the research, authorship, and/or publication of this article: This work was supported by JSPS KAKENHI Grant Numbers JP20K20624, JP21H03447, and JP22H03598; JST CREST Grant Number JPMJCR2112; and "Joint Usage/Research Center for Interdisciplinary Large-scale Information Infrastructures" in Japan (Project ID: jh230053-NAH, jh230009-NAHI).

## ORCID iDs

Qianxiang Ma  https://orcid.org/0000-0003-4688-5644
Rio Yokota  https://orcid.org/0000-0001-7573-7873

## Author biographies

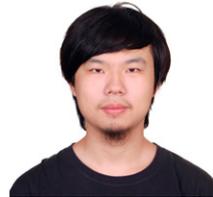

*Qianxiang Ma* is a Ph.D. student studying in the School of Computing, Tokyo Institute of Technology, under the guidance of Professor Rio Yokota. He received a B.S. in computer science from Case Western Reserve University, Cleveland, Ohio, United States, in 2018 and an M.Eng. in computer science from Tokyo Institute of Technology, Tokyo, Japan, in 2020. His current research topic includes optimizing and parallelizing the structured low-rank approximated matrices algorithms.

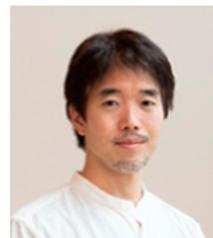

*Rio Yokota* is a Professor at the Global Scientific Information and Computing Center, Tokyo Institute of Technology. His research interests lie at the intersection of high performance computing, linear algebra, and machine learning. He is the developer numerous libraries for fast multipole methods (ExaFMM), hierarchical low-rank algorithms (Hatrix), and information matrices in deep learning (ASDL) that scale to the full system on the largest supercomputers today. He has been optimizing algorithms on GPUs since 2006, and was part of a team that received the Gordon Bell prize in 2009 using the first GPU supercomputer. Rio is a member of ACM, IEEE, and SIAM.